\pdfoutput=1
\documentclass[journal,twoside,web]{ieeecolor}
\input{settings/ieeecolor/tmi-template-settings.tex}

\usepackage{blindtext} %
\usepackage{mwe}  %

\usepackage[T1]{fontenc}  %
\usepackage[utf8]{inputenc}  %

\usepackage{newtxmath}
\usepackage{bm}  %
\DeclareMathAlphabet{\mathcal}{OMS}{cmsy}{m}{n}
\DeclareSymbolFont{AMSb}{U}{msb}{m}{n}
\DeclareSymbolFontAlphabet{\mathbb}{AMSb}

\usepackage{csquotes} %
\usepackage{textcomp}  %
\usepackage{siunitx}  %
\DeclareSIUnit[number-unit-product = {}]{\percent}{\char 37}
\DeclareSIUnit{\fps}{fps}
\usepackage{newunicodechar} %
\usepackage{pifont} %
\usepackage{bbding} %
\usepackage{xfrac}
\usepackage{nth} %

\usepackage[final, expansion=alltext, protrusion=alltext-nott]{microtype}
\DisableLigatures{encoding = T1, family = tt*}

\usepackage[english]{babel}
\addto\captionsenglish{%
}

\usepackage[style=base]{caption}  %
\usepackage{subcaption}
\DeclareCaptionLabelSeparator{ieeefigsep}{.\quad}
\DeclareCaptionLabelSeparator{ieeetmifigsep}{\textcolor{mblue}{.}\quad}
\usepackage{mathtools}  %
\usepackage{cases}

\usepackage{algorithm}
\usepackage{algpseudocode}
\usepackage[zerostyle=a,straightquotes]{newtxtt}

\usepackage{booktabs}
\usepackage{multirow}
\usepackage{threeparttable}
\makeatletter
\def\TPT@doparanotes{\par
   \prevdepth\z@ \TPT@hsize
   \TPTnoteSettings
   \parindent\z@ \pretolerance 8
   \linepenalty 200
   \renewcommand\item[1][]{\relax\ifhmode \begingroup
       \unskip
       \advance\hsize 10em %
       \penalty -45 \hskip\z@\@plus\hsize \penalty-19
       \hskip .15\hsize \penalty 9999 \hskip-.15\hsize
       \hskip .01\hsize\@plus-\hsize\@minus.01\hsize
       \hskip .5em\@plus .3em
      \endgroup\fi
      \tnote{##1}\ignorespaces}%
   \let\TPToverlap\relax
   \def\endtablenotes{\par}%
}
\makeatother
\renewcommand{\TPTnoteSettings}{%
    \setlength\leftmargin{1.5em}%
    \setlength\labelwidth{.5em}%
    \setlength\labelsep{0pt}%
    \rightskip\tabcolsep \leftskip\tabcolsep
}
\usepackage{tabularx}
\newcolumntype{L}[1]{>{\hsize=\if!#1!1\else#1\fi\hsize\raggedright\arraybackslash}X}
\newcolumntype{R}[1]{>{\hsize=\if!#1!1\else#1\fi\hsize\raggedleft\arraybackslash}X}
\newcolumntype{C}[1]{>{\hsize=\if!#1!1\else#1\fi\hsize\centering\arraybackslash}X}
\newcommand*{\mco}[1]{\multicolumn{1}{c}{#1}}

\usepackage{glossaries-extra}
\setabbreviationstyle[acronym]{long-short}  %

\usepackage{calc}
\usepackage{xifthen} %
\usepackage{etoolbox} %
\usepackage{xparse} %
\usepackage{xspace} %

\usepackage{soul}
\usepackage{printlen}

\usepackage{cite}

\usepackage{url}
\usepackage[bookmarks=false, hidelinks]{hyperref}
\usepackage[nameinlink, capitalize]{cleveref} %
\crefname{table}{Table}{Tables} %
\crefname{figure}{Fig.\@}{Fig.\@} %
\Crefname{figure}{Fig.\@}{Fig.\@} %
\crefname{section}{Section}{Sections} %
\Crefname{section}{Section}{Sections} %
\crefname{equation}{}{} %
\Crefname{equation}{Equation}{Equations} %
\crefname{algorithm}{Algorithm}{Algorithms} %
\Crefname{algorithm}{Algorithm}{Algorithms} %
\newcommand*{\figurenamecref}{Fig.}
\newcommand*{\figurenameplcref}{Fig.}
\crefformat{figure}{#2\color{mblue}\figurenamecref~#1#3}
\Crefformat{figure}{#2\color{mblue}\figurenamecref~#1#3}
\crefrangeformat{figure}{#3\color{mblue}\figurenameplcref~#1#4--#5\color{mblue}#2#6}
\Crefrangeformat{figure}{#3\color{mblue}\figurenameplcref~#1#4--#5\color{mblue}#2#6}
\crefmultiformat{figure}{#2\color{mblue}\figurenameplcref~#1#3}%
    { and~#2\color{mblue}#1#3}{, #2\color{mblue}#1#3}{, and~#2\color{mblue}#1#3}
\Crefmultiformat{figure}{#2\color{mblue}\figurenameplcref~#1#3}%
    { and~#2\color{mblue}#1#3}{, #2\color{mblue}#1#3}{, and~#2\color{mblue}#1#3}
\newcommand*{\tablenamecref}{Table}
\newcommand*{\tablenameplcref}{Tables}
\crefformat{table}{#2\color{mblue}\tablenamecref~#1#3}
\Crefformat{table}{#2\color{mblue}\tablenamecref~#1#3}
\crefrangeformat{table}{#3\color{mblue}\tablenameplcref~#1#4--#5\color{mblue}#2#6}
\Crefrangeformat{table}{#3\color{mblue}\tablenameplcref~#1#4--#5\color{mblue}#2#6}
\crefmultiformat{table}{#2\color{mblue}\tablenameplcref~#1#3}%
    { and~#2\color{mblue}#1#3}{, #2\color{mblue}#1#3}{, and~#2\color{mblue}#1#3}
\Crefmultiformat{table}{#2\color{mblue}\tablenameplcref~#1#3}%
    { and~#2\color{mblue}#1#3}{, #2\color{mblue}#1#3}{, and~#2\color{mblue}#1#3}

\let\originalleft\left
\let\originalright\right
\renewcommand{\left}{\mathopen{}\mathclose\bgroup\originalleft}
\renewcommand{\right}{\aftergroup\egroup\originalright}
\DeclarePairedDelimiter{\parens}{\lparen}{\rparen}
\DeclarePairedDelimiter{\braces}{\lbrace}{\rbrace}
\DeclarePairedDelimiter{\bracks}{\lbrack}{\rbrack}

\DeclarePairedDelimiter{\abs}{\lvert}{\rvert}

\DeclarePairedDelimiterXPP\onenorm[1]{}{\lVert}{\rVert}{_{1}}{#1}
\DeclarePairedDelimiterXPP\twonorm[1]{}{\lVert}{\rVert}{_{2}}{#1}
\DeclarePairedDelimiterXPP\fronorm[1]{}{\lVert}{\rVert}{_{\text{F}}}{#1}
\DeclarePairedDelimiterXPP\infnorm[1]{}{\lVert}{\rVert}{_{\infty}}{#1}
\DeclarePairedDelimiterXPP\pnorm[1]{}{\lVert}{\rVert}{_{p}}{#1}
\DeclarePairedDelimiterXPP\tvnorm[1]{}{\lVert}{\rVert}{_{\text{TV}}}{#1}
\DeclarePairedDelimiterXPP\maxnorm[1]{}{\lVert}{\rVert}{_{\text{max}}}{#1}
\DeclarePairedDelimiterXPP\nuclearnorm[1]{}{\lVert}{\rVert}{_{\ast}}{#1}

\DeclareMathOperator*{\conv}{\ast}

\newcommand*{\set}[1]{\mathbb{#1}}

\NewDocumentCommand{\realnumbers}{}{\set{R}}

\NewDocumentCommand{\ellipsis}{}{\dots}
\let\setroster\braces

\newcommand*{\logten}{\log_{10}}
\DeclareMathOperator{\sign}{sign}
\DeclareMathOperator{\sinc}{sinc}

\newcommand*{\mat}[1]{\bm{#1}}
\renewcommand*{\vec}[1]{\bm{#1}}

\newcommand*{\conj}{^*}
\newcommand*{\adj}{^*}

\DeclareMathOperator{\prox}{prox}

\DeclareMathOperator*{\argmin}{argmin}

\newcommand*{\estimate}[1]{\hat{#1}}
\newcommand*{\prediction}[1]{\hat{#1}}

\newcommand*{\expectedvalue}{\mathbb{E}}
\newcommand*{\variance}{\text{Var}}

\newcommand*{\dd}[1]{\mathop{\mathrm{d}#1}\nolimits}

\newcommand*{\dirac}{\delta}

\newcommand*{\argdot}{\,\cdot\,}

\newcommand*{\domsymb}{\Omega}
\newcommand*{\perturb}{\varepsilon}
\newcommand*{\baseindex}{i}

\newcommand*{\wavelength}{\lambda}

\newcommand*{\elemwidth}{d}
\newcommand*{\elemangle}{\theta}

\newcommand*{\rayleighpdfsymb}{f}
\newcommand*{\rayleighcdfsymb}{F}
\newcommand*{\rayleighquantilesumb}{Q}
\newcommand*{\rayleighpdf}{\rayleighpdfsymb}
\newcommand*{\rayleighcdf}{\rayleighcdfsymb}
\newcommand*{\rayleighquantile}{\rayleighquantilesumb}
\DeclareMathOperator{\rayleighdistsymb}{Rayleigh}
\newcommand*{\rayleighdist}{\rayleighdistsymb}

\newcommand*{\complexnormaldistsymb}{\mathcal{CN}}
\newcommand*{\complexnormaldist}{\complexnormaldistsymb}

\newcommand{\rndvaramplitude}{X}
\newcommand{\confidencelevel}{\beta}

\newcommand*{\imageletter}{x}
\newcommand*{\imcompindex}{k}
\newcommand*{\imcomponent}{\imageletter_{\imcompindex}}
\newcommand*{\imscl}{\imageletter}
\newcommand*{\imvec}{\vec{\imageletter}}

\newcommand*{\imvecapprox}{\tilde{\imvec}}
\newcommand{\impos}{\vec{r}}

\newcommand*{\imvecdim}{n}
\newcommand*{\imvecdom}{\realnumbers^{\imvecdim}}
\newcommand*{\imvecdef}{\imvec\in\imvecdom}
\newcommand*{\imvecgtsubspace}{V}
\newcommand*{\imvecapproxsubspace}{W}

\newcommand*{\rawdataletter}{y}
\newcommand*{\noiseletter}{n}
\newcommand*{\rawdata}{\rawdataletter}
\newcommand*{\rdvec}{\vec{\rawdataletter}}
\newcommand*{\rdvecdim}{m}
\newcommand*{\rdvecdom}{\realnumbers^{\rdvecdim}}
\newcommand*{\rdvecdef}{\rdvec\in\rdvecdom}
\newcommand*{\nvec}{\vec{\noiseletter}}
\newcommand*{\nvecdef}{\nvec\in\rdvecdom}

\newcommand*{\fieldpointletter}{r}
\newcommand*{\fieldpoint}{\vec{\fieldpointletter}}
\newcommand*{\fieldpointdom}{\domsymb}
\newcommand*{\trf}{\imageletter}
\newcommand*{\trfapprox}{\tilde{\imageletter}}
\newcommand*{\sirsymb}{h}

\newcommand*{\txsign}{\text{tx}}
\newcommand*{\rxsign}{\text{rx}}

\newcommand*{\rxnumber}{n_{r}}
\newcommand*{\txindex}{i}
\newcommand*{\rxindex}{j}
\newcommand*{\txsubscript}{i}
\newcommand*{\rxsubscript}{\mkern-3mu j}  %
\newcommand*{\txrxindex}{i, j}
\newcommand*{\sirtxspecifier}{^{\txsign}_{\txsubscript}}
\newcommand*{\sirrxspecifier}{^{\rxsign}_{\rxsubscript}}
\newcommand*{\sirtx}{\sirsymb\sirtxspecifier}
\newcommand*{\sirrx}{\sirsymb\sirrxspecifier}
\newcommand*{\sirapproxsymb}{\bar{\sirsymb}}

\newcommand*{\sirapproxtx}{\sirapproxsymb\sirtxspecifier}
\newcommand*{\sirapproxrx}{\sirapproxsymb\sirrxspecifier}
\newcommand*{\sirapproxtxpos}{\fieldpoint_{\txsubscript}}
\newcommand*{\sirapproxrxpos}{\fieldpoint_{\rxsubscript}}
\newcommand*{\pewaveformsymb}{v_{\text{pe}}}
\newcommand*{\pewaveform}{\pewaveformsymb}
\newcommand*{\delaysymb}{\tau}

\newcommand*{\delaytx}{\delaysymb\sirtxspecifier}
\newcommand*{\delayrx}{\delaysymb\sirrxspecifier}

\newcommand*{\dasopletter}{D}
\newcommand*{\dasopmat}{\mat{\dasopletter}}
\newcommand*{\dasopmatdef}{\dasopmat\colon\rdvecdom\to\imvecdom}
\newcommand*{\dasweightletter}{w}
\newcommand*{\dasweightopmat}{\mat{\MakeUppercase\dasweightletter}}
\newcommand*{\dasweight}{\dasweightletter}
\newcommand*{\dasweightmatopdef}{\dasweightopmat\colon\imvecdom\to\imvecdom}

\newcommand*{\physopletter}{H}
\newcommand*{\physopmat}{\mat{\physopletter}}
\newcommand*{\physopmatdef}{\physopmat\colon\imvecdom\to\rdvecdom}
\newcommand*{\adjphysopmat}{\mat{\physopletter} \adj}
\newcommand*{\adjphysopmatdef}{\adjphysopmat\colon\rdvecdom\to\imvecdom}
\newcommand*{\regularizer}{\mathcal{R}}
\newcommand*{\datafidelity}{\mathcal{D}}
\newcommand*{\regparam}{\varkappa}
\newcommand*{\gradientstep}{\gamma}
\newcommand*{\proxregparam}{\mu}
\newcommand*{\proxregexpl}{\prox_{\gradientstep\regparam\regularizer}}
\newcommand*{\iterindex}{k}
\newcommand*{\iter}[2]{#1^{\parens{#2}}}

\newcommand*{\cnnparams}{\vec{\theta}}
\newcommand*{\cnnletter}{f}
\newcommand*{\cnnsymb}{\vec{\cnnletter}_{\cnnparams}}
\newcommand*{\cnn}{\cnnsymb}
\newcommand*{\cnnchannumb}{N_c}

\newcommand*{\rescnnletter}{r}
\newcommand*{\rescnnsymb}{\vec{\rescnnletter}_{\cnnparams}}
\newcommand*{\rescnn}{\rescnnsymb}
\newcommand*{\cnndef}{\cnn\colon\imvecdom\to\imvecdom}

\newcommand*{\sltsymb}{g}
\newcommand*{\sltparam}{\alpha}
\newcommand*{\sltsf}{\sltsymb_{\sltparam}}

\newcommand*{\sltsfdef}{\sltsf\colon\realnumbers\to\realnumbers}

\newcommand*{\slt}{\sltsf}
\newcommand*{\sltdef}{\sltsfdef}

\newcommand*{\mutparam}{\mu}
\newcommand*{\mutsymb}{g}
\newcommand*{\mutsf}{\mutsymb_{\mutparam}}

\newcommand*{\mutsfdef}{\mutsf\colon\realnumbers\to\realnumbers}

\newcommand*{\mutransform}{\mutsf}
\newcommand*{\mutransformdef}{\mutsfdef}

\newcommand*{\empriskletter}{R}
\newcommand*{\emprisk}{\empriskletter}
\newcommand*{\losssymb}{\mathcal{L}}
\DeclareMathOperator{\mse}{\glsxtrshort{mse}}
\DeclareMathOperator{\mae}{\glsxtrshort{mae}}

\DeclareMathOperator{\mslae}{\glsxtrshort{mslae}}
\DeclareMathOperator{\mmuae}{\glsxtrshort{mmuae}}
\newcommand*{\loss}{\losssymb}
\newcommand*{\mseloss}{\losssymb_{\mse}}
\newcommand*{\maeloss}{\losssymb_{\mae}}
\newcommand*{\mslaeloss}{\losssymb_{\mslae}}
\newcommand*{\mmuaeloss}{\losssymb_{\mmuae}}

\newcommand*{\dsetindex}{\baseindex}
\newcommand*{\dsetsize}{l}
\newcommand*{\dsetpairs}{
    \setroster{
        \parens{ \iter{\imvec}{1}, \iter{\imvecapprox}{1} },
        \ellipsis{},
        \parens{ \iter{\imvec}{\dsetsize}, \iter{\imvecapprox}{\dsetsize} }
    }
}

\newcommand*{\ndim}[1]{\mbox{#1-D}}  %

\NewDocumentCommand{\margintodo}{ m o }{%
    \marginpar{\colorbox{yellow}{\parbox{\marginparwidth-\marginparsep}{\raggedright\scriptsize \textcolor{black}{#1}}}}%
    \IfValueT{#2}{\hl{#2}}%
}
\NewDocumentCommand{\inlinetodo}{ m o }{%
    \IfValueT{#2}{\hl{#2}}\raisebox{+0.85ex}{\tiny\hl{[#1]}}%
}

\newcommand{\invalidtabentry}{\( \times \)}
\newcommand{\tableheaderstyle}[1]{{#1}}
\newcommand{\tablefontstyle}{\sffamily}

\newcommand{\maxangleval}{29.7078}
\newcommand{\maxangle}{\theta_{r}}
\newcommand{\phtdomspecletter}{\Omega}
\newcommand{\phtdomimg}{\phtdomspecletter_{i}}
\newcommand{\phtdomred}{\phtdomspecletter_{r}}
\newcommand{\phtdomlim}{\phtdomspecletter_{l}}
\newcommand{\phtdomext}{\phtdomspecletter_{e}}

\newcommand{\sprctext}{ten} %
\newcommand{\sprcmmcubed}{153.35}
\newcommand{\scatterernumber}{900000}

\newcommand{\datasetsize}{31000}
\newcommand{\trainsetsize}{30000}
\newcommand{\validsetsize}{500}
\newcommand{\numtestsetsize}{300}

\newcommand{\stepsperepoch}{1000}
\newcommand{\iterationnumber}{500000}
\newcommand{\datasetlowervaldb}{-62}
\newcommand{\datasetuppervaldb}{+36}
\newcommand{\datasetrangevaldb}{98}
\newcommand{\datasetlowermeandb}{-50}
\newcommand{\datasetuppermeandb}{+30}

\newcommand{\datasetvalinterval}{90}

\newcommand*{\vsxtxfreq}{\SI{5.208}{\mega\hertz}}
\newcommand*{\vsxrxfreq}{\SI{20.833}{\mega\hertz}}

\newcommand*{\geaperture}{\SI{43.93}{\milli\meter}}
\newcommand*{\gecenterfreq}{\SI{5.3}{\mega\hertz}}
\newcommand*{\gebandwidth}{\SI{75}{\percent}}
\newcommand*{\geelemnumber}{\num{192}}
\newcommand*{\geelemwidth}{\SI{207}{\micro\meter}}
\newcommand*{\geelemheight}{\SI{6}{\milli\meter}}
\newcommand*{\geelevationfocus}{\SI{28}{\milli\meter}}
\newcommand*{\gepitch}{\SI{230}{\micro\meter}}

\newcommand{\meansoundspeedval}{1540}

\newcommand{\simrxfreq}{\SI{31.25}{\mega\hertz}}

\newcommand*{\imagesampling}{\wavelength{} / 4 \times{} \wavelength{} / 8}

\newcommand*{\numimdim}{\num{596 x 1600}}

\newcommand*{\numimdimpad}{\num{608 x 1600}}
\newcommand{\rescellx}{\num{0.71} \wavelength}
\newcommand{\rescelly}{\num{3.23} \wavelength}
\newcommand{\rescellz}{\num{1.10} \wavelength}
\newcommand{\rescell}{\rescellx \times \rescelly \times \rescellz}

\newcommand{\metricdominclsymb}{\text{I}}
\newcommand{\metricdomincl}{\domsymb_{\metricdominclsymb}}
\newcommand{\metricdomblocksymb}{\text{B}}
\newcommand{\metricdomblock}{\domsymb_{\metricdomblocksymb}}
\newcommand{\metricdomsl}{\domsymb_{\textnormal{\glsxtrshort{sl}}}}
\newcommand{\metricdomgl}{\domsymb_{\textnormal{\glsxtrshort{gl}}}}
\newcommand{\metricdomew}{\domsymb_{\textnormal{\glsxtrshort{ew}}}}
\newcommand{\metricdomlg}{\domsymb_{\textnormal{LG}}}
\newcommand{\metricdomsrsymb}{\text{S}}
\newcommand{\metricdomsr}{\domsymb_{\metricdomsrsymb}}
\newcommand{\metricreflectorA}{p_0}
\newcommand{\metricreflectorB}{p_1}
\newcommand{\metricreflectorC}{p_2}
\newcommand{\metricreflectorD}{p_3}
\newcommand{\metricexpdomcia}{\domsymb_{\textnormal{A}}}
\newcommand{\metricexpdomcib}{\domsymb_{\textnormal{B}}}
\newcommand{\metricexpdomcic}{\domsymb_{\textnormal{C}}}
\newcommand{\metricexpdombckg}{\domsymb_{\textnormal{D}}}
\newcommand{\metricexpdomsr}{\domsymb_{\textnormal{S}}}

\newcommand{\sectitlemethods}{Methods}
\newcommand{\sectitlehdrtraining}{Training on High-Dynamic-Range Data}
\newcommand{\sectitleexperiments}{Experiments}
\newcommand{\exptitletrainings}{Hyperparameter Search}
\newcommand{\exptitlesignaltype}{Image Representations}
\newcommand{\exptitlereferenceimage}{Reference Image Configurations}
\newcommand{\exptitleloss}{Training Losses}
\newcommand{\exptitlearchitecture}{Convolutional Blocks and Skip Connections}
\newcommand{\exptitlechannelnumber}{Initial Channel Expansion Numbers}
\newcommand{\exptitletrainingsize}{Training Set Sizes}
\newcommand{\exptitlekernelinitializer}{Kernel Initializers}
\newcommand{\exptitlelearningrates}{Learning Rates}
\newcommand{\exptitlenumerical}{Numerical Test Phantom}
\newcommand{\exptitleexperimental}{Experimental Evaluations}

\newcommand{\resnumcnnAlgd}{\glsxtrshort{mse}-\num{16}}
\newcommand{\resnumcnnBlgd}{\glsxtrshort{mae}-\num{16}}
\newcommand{\resnumcnnClgd}{\glsxtrshort{mslae}-\num{16}}
\newcommand{\resnumcnnDlgd}{\glsxtrshort{mslae}-\num{32}}
\newcommand{\resexpcnnAlgd}{\resnumcnnClgd}

\newcommand{\mslaelosseq}{%
    \mslaeloss\parens{\imvec, \prediction{\imvec}}
    =
    \frac{1}{\imvecdim}
    \onenorm{ \slt\parens{\imvec} - \slt\parens{\prediction{\imvec}} }
}
\newcommand{\sltransformeq}{%
    \sltsf\parens{\imcomponent}
    =
    \sign\parens{\imcomponent}
    \log_{\sltparam}\parens*{
        \frac{
            \sltparam
        }{
            \max\parens{\sltparam, \abs{\imcomponent}}
        }
    }
}

\newcommand*{\tfver}{v1.14}
\newcommand*{\trtver}{v5.1.5}
\newcommand*{\tfurl}{\url{https://www.tensorflow.org}}
\newcommand*{\trturl}{\url{https://developer.nvidia.com/tensorrt}}
\newcommand*{\pyusurl}{\url{https://gitlab.com/pyus/pyus}}

\newglossaryentry{python}{%
    name={Python},
    description={interpreted, high-level, general-purpose programming language}
}
\newglossaryentry{matlab}{%
    name={MATLAB},
    description={multi-paradigm numerical computing environment and proprietary programming language developed by MathWorks}
}
\newglossaryentry{tensorflow}{%
   name={TensorFlow},
   description={An end-to-end open source machine learning platform}
}
\newglossaryentry{tensorrt}{%
   name={TensorRT},
   description={NVIDIA TensorRT™ is a platform for high-performance deep learning inference.}
}
\newglossaryentry{mu-law}{%
    name={\textmu-law},
    description={Comanding algorithm}
}
\newglossaryentry{ge}{%
    name={General Electric Company},
    description={TODO}
}

\newglossaryentry{ge-healthcare}{%
    name={GE Healthcare},
    description={TODO}
}
\newcommand{\gehealthcareaddress}{Chicago, IL, USA}
\newglossaryentry{ge9ld}{%
    name={9L-D},
    description={GE Linear transducer}
}
\newglossaryentry{bmode}{%
    name={B-mode},
    description={Brightness mode}
}
\newglossaryentry{verasonics}{%
    name={Verasonics},
    description={TODO}
}
\newglossaryentry{vantage256}{%
    name={Vantage~256},
    description={TODO}
}
\newcommand{\verasconicsaddress}{Kirkland, WA, USA}
\newglossaryentry{cirs}{%
    name={CIRS},
    description={Multi-purpose multi-tissue ultrasound phantom}
}
\newcommand{\cirsaddress}{Norfolk, VA, USA}
\newglossaryentry{cirs-040gse}{%
    name={\gls{cirs} model 040GSE},
    description={Multi-purpose multi-tissue ultrasound phantom}
}
\newglossaryentry{cirs-054gs}{%
    name={\gls{cirs} model 054GS},
    description={General-purpose ultrasound phantom}
}
\newglossaryentry{pyus}{%
    name={PyUS},
    description={TODO}
}
\newglossaryentry{fieldii}{%
    name={Field~II},
    description={TODO}
}
\newglossaryentry{x-ray}{%
    name={X-ray},
    description={X-radiation or Röntgen radiation}
}
\newglossaryentry{nvidia}{%
    name={NVIDIA},
    description={American technology company incorporated in Delaware and based in Santa Clara, California. It designs graphics processing units (GPUs) for the gaming and professional markets, as well as system on a chip units (SoCs) for the mobile computing and automotive market.}
}
\newglossaryentry{1080ti}{
    name={NVIDIA GeForce GTX 1080~Ti},
    description={Specific GPU from NVIDIA (Pascal architecture)}
}
\newglossaryentry{mx150}{
    name={NVIDIA GeForce MX~150},
    description={Specific GPU from NVIDIA (Pascal architecture)}
}
\newglossaryentry{titanv}{
    name={NVIDIA TITAN~V},
    description={Specific GPU from NVIDIA (Volta architecture)}
}
\newglossaryentry{v100}{
    name={NVIDIA Tesla V100},
    description={Specific GPU from NVIDIA (Volta architecture)}
}
\newglossaryentry{unet}{
    name={U-Net},
    description={TODO}
}
\newglossaryentry{resnet}{
    name={ResNet},
    description={Typical neural network architecture}
}
\newglossaryentry{he}{
    name={He},
    description={TODO He init}
}
\newglossaryentry{glorot}{
    name={Glorot},
    description={TODO Gloro init}
}
\newglossaryentry{bspline}{
    name={B-spline},
    description={TODO}
}
\newglossaryentry{etc}{
    name={etc.},
    description={et cetera},
}
\newglossaryentry{ie}{
    name={i.e.}, description={id est}, %
}
\newglossaryentry{eg}{
    name={e.g.}, description={exempli gratia}, %
}
\newglossaryentry{etal}{
    name={\textit{et al.}}, description={et alii}, %
}
\newglossaryentry{ibid}{
    name={\textit{ibid.}},
    description={ibidem}, %
}
\newglossaryentry{vs}{
    name={vs.}, description={TODO}, %
}
\newglossaryentry{wrt}{
    name={w.r.t.}, description={TODO}, %
}
\newglossaryentry{esp}{
    name={esp.}, description={TODO}, %
}
\newglossaryentry{invitro}{
    name={\textit{in vitro}}, description={TODO}, %
}
\newglossaryentry{invivo}{
    name={\textit{in vivo}}, description={TODO}, %
}
\newglossaryentry{exvivo}{
    name={\textit{ex vivo}}, description={TODO}, %
}
\newglossaryentry{insitu}{
    name={\textit{in situ}}, description={TODO}, %
}
\newglossaryentry{insilico}{
    name={\textit{in silico}}, description={TODO}, %
}
\newglossaryentry{interalia}{
    name={\textit{inter alia}}, description={TODO}, %
}
\newglossaryentry{intoto}{
    name={\textit{in toto}}, description={TODO}, %
}
\newglossaryentry{apriori}{
    name={\textit{a priori}}, description={TODO}, %
}
\newglossaryentry{aposteriori}{
    name={\textit{a posteriori}}, description={TODO}, %
}
\newglossaryentry{alas}{
    name={\textit{alas}}, description={TODO}, %
}
\newglossaryentry{defacto}{
    name={de facto}, description={TODO}, %
}

\newacronym{dl}{DL}{deep learning}
\newacronym{nn}{NN}{neural network}
\newacronym{dnn}{DNN}{deep neural network}
\newacronym{cnn}{CNN}{convolutional neural network}
\newacronym{relu}{ReLU}{rectified linear unit}
\newacronym{conv-layer}{CL}{convolutional layer}
\newacronym{std-conv-block}{FCB}{fully convolutional block}
\newacronym{res-conv-block}{RCB}{residual convolutional block}
\newacronym{mse}{MSE}{mean squared error}
\newacronym{mae}{MAE}{mean absolute error}
\newacronym{sdbae}{MSDBAE}{mean signed dB absolute error}
\newacronym{mslae}{MSLAE}{mean signed logarithmic absolute error}
\newacronym{mmuae}{MMUAE}{mean \textmu-law absolute error}
\newacronym[longplural={regions of interest}]{roi}{ROI}{region of interest}
\newacronym{voi}{VOI}{volume of interest}
\newacronym{fps}{FPS}{frames per second}
\newacronym{snr}{SNR}{signal-to-noise ratio}
\newacronym{cnr}{CNR}{contrast-to-noise ratio}
\newacronym{gcnr}{GCNR}{generalized contrast-to-noise ratio}
\newacronym{fwhm}{FWHM}{full width at half maximum}
\newacronym{ssim}{SSIM}{structural similarity}
\newacronym{psnr}{PSNR}{peak signal-to-noise ratio}
\newacronym{dra}{DRA}{dynamic range alteration}
\newacronym{mri}{MRI}{magnetic resonance imaging}
\newacronym{ct}{CT}{computed tomography}
\newacronym{hdr}{HDR}{high dynamic range}
\newacronym{nurbs}{NURBS}{non-uniform rational B-spline}
\newacronym{fft}{FFT}{fast Fourier transform}
\newacronym{ifft}{IFFT}{inverse fast Fourier transform}
\newacronym{bp}{BP}{backprojection}
\newacronym{fbp}{FBP}{filtered backprojection}
\newacronym{us}{US}{ultrasound}
\newacronym{das}{DAS}{delay-and-sum}
\newacronym{trf}{TRF}{tissue reflectivity function}
\newacronym{pw}{PW}{plane wave}
\newacronym{cpwc}{CPWC}{coherent plane-wave compounding}
\newacronym{sa}{SA}{synthetic aperture}
\newacronym{dw}{DW}{diverging wave}
\newacronym{rf}{RF}{radio frequency}
\newacronym{prf}{PRF}{pulse repetition frequency}
\newacronym{tgc}{TGC}{time gain compensation}
\newacronym{lq}{LQ}{low-quality}
\newacronym{hq}{HQ}{high-quality}
\newacronym{uq}{UQ}{ultrahigh-quality}
\newacronym{picmus}{PICMUS}{plane-wave imaging challenge in medical ultrasound}
\newacronym{sir}{SIR}{spatial impulse response}
\newacronym{psf}{PSF}{point spread function}
\newacronym{iq}{IQ}{in-phase quadrature}
\newacronym{gl}{GL}{grating lobe}
\newacronym{sl}{SL}{side lobe}
\newacronym{ew}{EW}{edge wave}
\newacronym{tx}{Tx}{transmit}
\newacronym{rx}{Rx}{receive}
\newacronym{mv}{MV}{minimum variance}
\newacronym{mi}{MI}{mechanical index}
\newacronym{piv}{PIV}{particle image velocimetry}
\newacronym{pdf}{PDF}{probability density function}
\newacronym{cdf}{CDF}{cumulative distribution function}
\newacronym{wss}{WSS}{wide-sense stationary}
\newacronym{acf}{ACF}{autocovariance function}
\newacronym{pcc}{PCC}{Pearson correlation coefficient}
\newacronym{epfl}{EPFL}{École polytechnique fédérale de Lausanne}
\newacronym{lts5}{LTS5}{Signal Processing Laboratory 5}
\newacronym{chuv}{CHUV}{University Hospital Center}
\newacronym{unil}{UNIL}{University of Lausanne}
\newacronym{cibm}{CIBM}{Center for Biomedical Imaging}
\newacronym{gpu}{GPU}{graphics processing unit}

\newcommand{\papertitle}{%
    CNN-Based Image Reconstruction Method for Ultrafast Ultrasound Imaging
}
\title{\papertitle}

\newcommand*{\correspondingemail}{dimitris.perdios@epfl.ch}
\newcommand*{\correspondingauthor}{Dimitris Perdios}

\author{%
    Dimitris~Perdios,~\IEEEmembership{Member,~IEEE,}
    Manuel~Vonlanthen,
    Florian~Martinez,~\IEEEmembership{Member,~IEEE,}
    Marcel~Arditi,~\IEEEmembership{Senior~Member,~IEEE,}
    and
    Jean-Philippe~Thiran,~\IEEEmembership{Senior~Member,~IEEE}%
    \thanks{%
        This work was supported in part by the Swiss National Science Foundation
        under Grant 205320\_175974 and Grant 206021\_170758.
        \textit{%
            (Corresponding author:
            \href{mailto:\correspondingemail}{\correspondingauthor}.)
        }
    }%
    \thanks{%
        Dimitris Perdios, Manuel Vonlanthen, Florian Martinez, and Marcel Arditi
        are with the
        \glsxtrfull{lts5},
        \glsxtrfull{epfl},
        1015 Lausanne,
        Switzerland
        (email: \href{mailto:\correspondingemail}{\correspondingemail}).
    }%
    \thanks{%
        Jean-Philippe Thiran
        is with the
        \glsxtrfull{lts5},
        \glsxtrfull{epfl},
        1015 Lausanne,
        Switzerland,
        also with the
        Department of Radiology,
        \glsxtrfull{chuv}
        and
        \glsxtrfull{unil},
        1011 Lausanne,
        Switzerland,
        and also with the
        CIBM Center for Biomedical Imaging,
        1015 Lausanne,
        Switzerland
        (email:
        \href{mailto:jean-philippe.thiran@epfl.ch}{jean-philippe.thiran@epfl.ch}%
        ).
    }%
    \thanks{%
        This article has supplementary material provided by the authors.
        Data is available online at \url{https://dx.doi.org/10.21227/vn0e-cw64}.
        Code is available online at \url{https://github.com/dperdios/dui-ultrafast}.
    }
}

\makeatletter
    \let\@savedtitle\@maketitle
    \let\savedmaketitle\maketitle
    \let\savedthanks\thanks
    \newcommand{\makesuptitle}{%
        \let\currenttitle\@title
        \newcommand{\suptitle}{\currenttitle{}\\\textit{Supplementary Material}}
        \title{\suptitle}
        \let\@maketitle\@savedtitle
        \let\maketitle\savedmaketitle
        \let\thanks\savedthanks  %
        \maketitle
    }
\makeatother

\newcommand{\supplements}{%
    \clearpage
    \setcounter{page}{1}%
    \setcounter{section}{0}%
    \setcounter{subsection}{0}%
    \setcounter{subsubsection}{0}%
    \setcounter{paragraph}{0}%
    \setcounter{table}{0}%
    \setcounter{figure}{0}%
    \setcounter{equation}{0}%
    \renewcommand{\theHsection}{Supplement.\thesection}%
    \renewcommand{\theHtable}{Supplement.\thetable}%
    \renewcommand{\theHfigure}{Supplement.\thefigure}%
    \renewcommand{\theHequation}{Supplement.\theequation}%

    \newcommand{\thesupplementprefix}{S}
    \renewcommand{\thesection}{\thesupplementprefix-\Roman{section}}%
    \renewcommand{\thetable}{\thesupplementprefix-\Roman{table}}%
    \renewcommand{\thefigure}{\thesupplementprefix\arabic{figure}}%
    \renewcommand{\theequation}{\thesupplementprefix\arabic{equation}}%

    \makesuptitle{}
}

\begin{document}

\maketitle

\glsresetall{}

\begin{abstract}
Ultrafast \gls{us} revolutionized biomedical imaging
with its capability of acquiring full-view frames at over \SI{1}{\kilo\hertz},
unlocking breakthrough modalities such as shear-wave
elastography and functional \gls{us} neuroimaging.
Yet,
it suffers from strong diffraction artifacts,
mainly caused by \glsxtrlongpl{gl}, \glsxtrlongpl{sl}, or \glsxtrlongpl{ew}.
Multiple acquisitions are typically required to obtain a sufficient
image quality,
at the cost of a reduced frame rate.
To answer the increasing demand for high-quality imaging
from single unfocused acquisitions,
we propose a two-step \gls{cnn}-based image reconstruction method,
compatible with real-time imaging.
A low-quality estimate is obtained by means of a backprojection-based operation,
akin to conventional \glsxtrlong{das} beamforming,
from which a high-quality image is restored using a residual \gls{cnn}
with multiscale and multichannel filtering properties,
trained specifically to remove the diffraction artifacts
inherent to ultrafast \gls{us} imaging.
To account for both the \glsxtrlong{hdr} and the oscillating properties
of \glsxtrlong{rf} \gls{us} images,
we introduce the \gls{mslae} as a training loss function.
Experiments were conducted with a linear transducer array,
in single plane-wave (\glsunset{pw}\gls{pw}) imaging.
Trainings were performed on a simulated dataset,
crafted to contain a wide diversity of structures and echogenicities.
Extensive numerical evaluations demonstrate that the proposed approach
can reconstruct images from single \glspl{pw} with a quality similar to
that of gold-standard \glsxtrlong{sa} imaging,
on a dynamic range in excess of \SI{60}{\decibel}.
\Gls{invitro} and \gls{invivo} experiments show that trainings carried out
on simulated data perform well in experimental settings.
\end{abstract}

\begin{IEEEkeywords}
\Glsxtrfullpl{cnn},
deep learning,
diffraction artifacts,
\glsxtrfull{hdr},
image reconstruction,
image restoration,
ultrafast \glsxtrfull{us} imaging.
\end{IEEEkeywords}

\glsresetall{}

\vfill\break
\section{Introduction}%
\label{sec:introduction}

\glsunset{us}  %
\IEEEPARstart{U}{ltrasound} (\glsxtrshort{us}) imaging is one
of the most widely used medical imaging modalities,
thanks to being non-ionizing,
and having a greater cost-effectiveness and portability
compared with \gls{x-ray} \gls{ct} or \gls{mri}.
Pulse-echo \gls{us} imaging is typically performed
by transmitting short acoustic pulses through a medium of interest
using an array of transducer elements,
and receiving echoes backscattered
from local variations in acoustic impedance.
Compared with conventional line-by-line scanning,
where sequential pulse-echo acquisitions are performed
using focused transmit beams for each image scan line,
ultrafast \gls{us} imaging relies on the insonification of the entire
field of view at once by transmitting a single unfocused wavefront,
such as a \gls{pw} or a \gls{dw}.
This strategy allows for extremely high frame rates of multiple
kilohertz~\cite{Tanter_UFFC_2014},
limited only by the round-trip time-of-flight of the transmitted wavefront.
Coupled with advances in electronics and software-based \gls{das} beamforming,
ultrafast \gls{us} imaging unlocked,
in the past two decades,
breakthrough imaging modalities such as
shear-wave elastography~\cite{Bercoff_UFFC_2004},
functional \gls{us} neuroimaging~\cite{Mace_NMETH_2011},
ultrasensitive \ndim{2} motion estimation~\cite{Tanter_UFFC_2002},
and high frame-rate vector flow imaging~\cite{Udesen_UFFC_2008}.

The main disadvantage of ultrafast \gls{us} imaging using single unfocused
transmit wavefronts is a decrease in image quality.
Indeed,
compared with a focused transmit beam which concentrates most of its energy
in a limited \gls{roi},
the energy of an unfocused wavefront is spread over the entire field of view,
resulting in backscattered echoes of lower amplitude and measurements
with lower \gls{snr}.
The absence of transmit focusing also results in a broader main lobe of
the \gls{psf},
consequently degrading the image resolution.
Furthermore,
diffraction artifacts,
such as the ones caused by \glspl{gl}, \glspl{sl}, and \glspl{ew},
are more pronounced in ultrafast \gls{us} imaging.
These artifacts can hamper lesion detectability
and displacement estimates~\cite{Montaldo_UFFC_2009},
especially when imaging highly heterogeneous tissue.

A common strategy to increase image quality in ultrafast \gls{us} imaging
consists of coherently compounding low-quality images obtained from multiple,
differently steered,
unfocused transmit wavefronts~\cite{Montaldo_UFFC_2009,%
Papadacci_UFFC_2014,Tanter_UFFC_2014,JensenJonas_UFFC_2016}.
While this method successfully improves the image quality
by increasing the number of steered acquisitions,
it inevitably comes at the cost of lower frame rates,
larger data transfers,
and increased computational requirements,
as multiple transmit-receive events and image reconstruction processes
are required.
\Gls{sa} imaging is another acquisition strategy based on
the coherent compounding of multiple low-quality images,
for which each transducer element is used in sequence to transmit
a wide \gls{dw}~\cite{Jensen_ULTRAS_2006}.
As sequential transmit-receive acquisition events
are required for compounding methods,
they are also subject to potential inter-acquisition tissue motion,
which results in blurring artifacts~\cite{Denarie_TMI_2013}.

Coherent compounding techniques may not be deployable in some applications,
such as those with extreme displacement estimation constraints
or low-power requirements,
where only a minimum number of transmit-receive events may be performed.
This implies a need for image reconstruction methods capable of extracting
more information from ultrafast acquisitions,
which gave rise to the \gls{picmus}~\cite{Liebgott_IUS_2016}.
Promising results were obtained using regularization techniques,
such as elastic net~\cite{Byram_UFFC_2015},
sparsity in wavelet bases~\cite{Besson_UFFC_2018},
or a sum of multiple regularizers~\cite{Ozkan_UFFC_2018}.
However,
as opposed to other imaging techniques
which can rely on robust regularizers (\gls{eg}, \gls{ct}),
common regularizers are not well suited to the statistical
properties of \gls{us} images,
especially in the presence of speckle patterns.
This imposes an image-dependent fine-tuning of hyperparameters,
limiting the appeal of these approaches except in specific cases.

Deep learning entered the medical
image analysis field~\cite{Greenspan_TMI_2016},
quickly followed by the image reconstruction
one~\cite{Wang_ACCESS_2016,Wang_TMI_2018},
with tight links to inverse
problems~\cite{Jin_TIP_2017,McCann_SPM_2017,Gupta_TMI_2018,Lucas_SPM_2018}.
As \gls{us} imaging is achieved through a sophisticated
signal processing pipeline,
deep learning-based components may be introduced
at many stages of this process~\cite{vanSloun_JPROC_2020}.
Different strategies relying on \glspl{cnn} have been proposed
for post-beamforming
speckle reduction~\cite{Vedula_ARXIV_2017,Dietrichson_IUS_2018,Hyun_UFFC_2019}
or for mimicking the post-processing of clinical scanners~\cite{Huang_TMI_2020}.
Fully connected neural networks operating on beamformed signals
were proposed to remove off-axis artifacts~\cite{Luchies_TMI_2018}
or to learn the apodization weights of an adaptive
beamformer~\cite{Luijten_TMI_2020}.
An end-to-end \gls{cnn}-based method was proposed for segmenting anechoic cysts
from raw element data directly~\cite{Nair_UFFC_2020}.
Restoration techniques using \glspl{cnn} were proposed to enhance low-quality
images~\cite{Perdios_IUS_2018,Mishra_SPL_2019},
to learn a compounding operation from a reduced
number of insonifications~\cite{Gasse_UFFC_2017,Zhou_UFFC_2018},
or for super-resolution in the context of \gls{us} localization
microscopy~\cite{vanSloun_ICASSP_2019}.

The main objective of this work is to provide images with a minimum amount
of diffraction artifacts at the highest possible frame rate.
Inspired by regularized regression methods and~\cite{Jin_TIP_2017},
we propose a \enquote{two-step} image reconstruction approach,
consisting of a backprojection operation to obtain a low-quality image estimate,
followed by applying a \gls{cnn} trained to restore a high-quality image.
The backprojection operator is derived from linear acoustics and far-field
assumptions~\cite{Besson_UFFC_2018,Besson_TCI_2019},
resulting in an operation similar to \gls{das} beamforming,
and is further improved with a re-weighting operation.
The \gls{cnn} architecture is based on~\cite{Jin_TIP_2017} and~\cite{Gupta_TMI_2018},
with notable improvements over our preliminary work~\cite{Perdios_IUS_2018}.
To account for the \gls{hdr} property of US images
while preserving their \gls{rf} nature,
we introduce the \gls{mslae} as a training loss function.
Experiments were conducted on a linear transducer array using a single
\gls{pw} with normal incidence to reconstruct low-quality input images.
Reference images were reconstructed from the complete set of \gls{sa}
acquisitions.
The training of the \gls{cnn} was performed using a simulated-image dataset
with relevant statistical properties,
in particular spanning a wide dynamic range.
Special attention was given to speckle patterns
as they are essential to most displacement estimation techniques
deployed in ultrafast \gls{us} applications~\cite{Tanter_UFFC_2014,%
Bercoff_UFFC_2004,Mace_NMETH_2011,Tanter_UFFC_2002,Udesen_UFFC_2008}.
Extensive quantitative evaluations were performed on a numerical test phantom
inspired by~\cite{Rindal_UFFC_2019},
and robustness to experimental data was evaluated in both
\gls{invitro} and \gls{invivo} settings.
Current limitations and directions for future improvements are discussed.
Supplementary Material is also provided,
including in-depth statistical and mathematical derivations
of concepts involved in the proposed method,
extensive experiments performed for hyperparameter search,
and additional results obtained.
Data and code are available online (see first footnote).

\section{\sectitlemethods}%
\label{sec:methods}

\subsection{Background on Pulse-Echo Modeling and Imaging}%
\label{sec:methods:theory}

In this section,
we briefly summarize the \gls{sir} model~\cite{Jensen_JASA_1991}
that was used to generate a realistic training dataset.
We introduce the additional approximations made to obtain a computationally
tractable measurement model (forward operator),
which defines the inverse problem considered,
and whose adjoint (backprojection operator)
forms the basis of the proposed \gls{cnn}-based image reconstruction method
(\cref{sec:methods:overview}).
Brief notes on regularized regression techniques that served as inspiration
are also provided.

Under the first-order Born approximation,
assuming (longitudinal) linear acoustics,
and neglecting dispersive attenuation,
the signal received (\gls{eg}, by a transducer element)
from a weakly scattering medium \( \fieldpointdom \)
embedded in a homogeneous medium
and insonified by a transmitter (\gls{eg}, wavefront)
can be (compactly) expressed
as~\cite{Jensen_JASA_1991}
\begin{align}
    \rawdata_{\txrxindex}\parens{t}
    =
    \pewaveform\parens{t}
    \conv_{t}
    \int\limits_{\fieldpoint \in  \fieldpointdom}
    \bracks*{
        \sirtx\parens{\fieldpoint, t} \conv_{t} \sirrx\parens{\fieldpoint, t}
    }
    \trf\parens{\fieldpoint}
    \dd{\fieldpoint}
    ,
    \label{eq:sir-pe-integral-model}
\end{align}
where \( \conv_{t} \) denotes time convolution.
The terms \( \sirtx \) and \( \sirrx \) represent the \glspl{sir}
of the transmitter and receiver, respectively.
The pulse-echo waveform%
\footnote{Originally called pulse-echo wavelet in~\cite{Jensen_JASA_1991}.}
\( \pewaveform \) includes both electro-acoustic (transmit) and
acousto-electric (receive) impulse responses
as well as the electric excitation waveform
(assumed identical for all transducer elements).
Local fluctuations in density and propagation velocity,
which induce scattered echo signals,
are represented by \( \trf \).

Due to the high complexity of the \glspl{sir} involved
in~\cref{eq:sir-pe-integral-model},
and in order to obtain a computationally tractable measurement model
for use in image reconstruction methods,
further assumptions are commonly made~\cite{Besson_UFFC_2018, Besson_TCI_2019}.
Assuming far-field approximation both
for the transmitter (\gls{eg}, ideal wavefront)
and for the receiver (\gls{eg}, narrow transducer element),
and assuming \( \pewaveform \) to be a Dirac delta function \( \dirac \),
\cref{eq:sir-pe-integral-model} can be approximated as
\begin{align}
    \rawdata_{\txrxindex}\parens{t}
    =
    \int\limits_{\fieldpoint \in  \fieldpointdom}
    \sirapproxtx\parens{\fieldpoint}
    \sirapproxrx\parens{\fieldpoint}
    \dirac(
        t - \delaytx\parens{\fieldpoint} - \delayrx\parens{\fieldpoint}
    )
    \trf\parens{\fieldpoint}
    \dd{\fieldpoint}
    ,
    \label{eq:physical-operator}
\end{align}
where \( \sirapproxtx \) and \( \sirapproxrx \)
are scalar functions representing the (far-field) acoustic diffraction effects
of the transmitter and the receiver
to and from a field point \( \fieldpoint \),
respectively.
The terms \( \delaytx \) and \( \delayrx \)
represent the acoustic wave propagation times
from the transmitter and the receiver
to a field point \( \fieldpoint \),
respectively.

\Glsxtrlong{us} transducers typically comprise a set of \( \rxnumber \)
receivers (\gls{ie}, transducer elements) arranged in a regular array
and measurements are sampled at discrete time intervals.
As images are composed of discrete (pixel) values,
the (approximated) linear physical measurement model
defined in~\cref{eq:physical-operator}
can be conveniently expressed as a discretized operation
for all receivers as
\( \rdvec = \physopmat \imvec + \nvec \),
where \( \physopmatdef \) is the measurement (matrix) operator,
\( \imvecdef \) is the (vectorized) image we seek to recover,
\( \rdvecdef \) are the (vectorized) transducer-element measurements
(raw data),
and \( \nvecdef \) is the measurement noise.
(To lighten notations, the transmitter index \( \txindex \) has been omitted
in the matrix expressions.)
Recovering \( \imvec \) from \( \rdvec \) is a well-known
inverse problem in medical imaging and may be addressed
using various strategies~\cite{Gupta_TMI_2018}.

Classical backprojection algorithms,
which exploit the adjoint operator
\( \adjphysopmatdef \),
may be used to address such an inverse problem.
Using~\cref{eq:physical-operator}, one can express the adjoint
operation for all receivers
in the continuous domain as~\cite{Besson_TCI_2019}
\begin{align}
    \trfapprox_{\txindex}\parens{\fieldpoint}
    =
    \sirapproxtx\parens{\fieldpoint}
    \sum_{\rxindex=1}^{\rxnumber}
    \sirapproxrx\parens{\fieldpoint}
    \rawdata_{\txrxindex}\parens{
        \delaytx\parens{\fieldpoint} + \delayrx\parens{\fieldpoint}
    }
    ,
    \label{eq:physical-operator-adjoint}
\end{align}
which is an instance of the well-known \gls{das} algorithm,
where \( \sirapproxtx \) and \( \sirapproxrx \) can be interpreted as
weighting functions
(sometimes referred to as apodization functions).
As the resulting \gls{rf} image \( \trfapprox_{\txsubscript} \)
is obtained from a single insonification (transmitter),
it usually suffers from strong diffraction artifacts.
To improve the image quality,
it is common to average coherently multiple (low-quality) \gls{rf} images
reconstructed from different insonifications,
for instance using steered \glspl{pw} or \glspl{dw}
in a process called coherent compounding~\cite{Tanter_UFFC_2002},
or using \gls{sa} imaging~\cite{Jensen_ULTRAS_2006}.

As an alternative to using multiple insonifications,
regularization techniques may be deployed~\cite{Besson_UFFC_2018}.
Such methods involve iterative algorithms that rely on
backprojection-based image estimates and \enquote{denoising} projections
based on some regularizer used to infer prior knowledge
on the expected image~\cite{Combettes_BOOK_2011}.
These methods remain seldom used in \gls{us} imaging given the difficulty
of finding a regularizer suitable for the very specific and broad statistical
properties of \gls{us} images
(especially with speckle patterns).
Yet,
the strategy of applying \enquote{denoising} projections
to backprojection-based estimates is the basis of the proposed method.
(Additional details and insights are provided in the Supplementary Material,
\cref{sec:sup:methods:regularization}.)

\subsection{Proposed Image Reconstruction Method}%
\label{sec:methods:overview}

The proposed method consists of first computing a low-quality estimate
of \( \imvec \), denoted \( \imvecapprox \),
from measurements \( \rdvec \) acquired with a single insonification,
by means of a re-weighted backprojection-based \gls{das}
algorithm \( \dasopmatdef \).
We define \( \dasopmat \coloneq \dasweightopmat \adjphysopmat \),
where \( \dasweightmatopdef \)
is a \enquote{pixel-wise} re-weighting operator (diagonal matrix) defined,
for the \( i \)th transmitter,
as
\begin{align}
    \dasweight_{\txindex}\parens{\fieldpoint}
    =
    \parens[\Bigg]{
        \sirapproxtx\parens{\fieldpoint}
        \sum_{\rxindex=1}^{\rxnumber}
        \sirapproxrx\parens{\fieldpoint}
    }^{-1}
    .
    \label{eq:das-reweighting}
\end{align}
It has been designed to compensate for the amplitude-related effects
of (far-field) diffraction and can be interpreted
as post-\gls{das} image equalization.
In a second step, the resulting approximation
\( \imvecapprox = \dasopmat \rdvec \)
is fed to a \gls{cnn} \( \cnndef \),
with parameters \( \cnnparams \),
trained to recover a high-quality estimate of \( \imvec \) as
\( \estimate{\imvec} = \cnn\parens{\imvecapprox} \).
As opposed to end-to-end approaches seeking to map
a measurement space to an image space directly,
we suggest using a \gls{cnn} for a task in which they are well-known to excel,
namely restoration tasks~\cite{McCann_SPM_2017,Weigert_NMETH_2018}.

Put formally, we seek to train a mapping \( \cnn \) between a subspace
of low-quality images
\( \imvecapproxsubspace \subset \imvecdom \)
to a subspace of \enquote{ground-truth} images
\( \imvecgtsubspace \subset \imvecdom \).
To define these subspaces more precisely,
let us consider an \gls{us} transducer with a given aperture,
composed of an array of transducer elements,
with given geometry, center frequency, and bandwidth properties,
and designed to operate at a given transmit frequency.
We define \( \imvecapproxsubspace \) as the subspace of
\gls{us} images reconstructed using \( \dasopmat \) from measurements
acquired by a single insonification using the entire aperture.
These images are typically contaminated by high \gls{sl} and \gls{ew} artifacts
as well as potential \gls{gl} artifacts,
if the spatial sampling of the aperture is suboptimal
(\gls{eg}, linear-array designs).
For \( \imvecgtsubspace \),
we propose to use a transducer similar to the one used for
\( \imvecapproxsubspace \),
namely spanning the same aperture and
composed of transducer elements with the same physical properties,
but with a spatial sampling ensuring the absence of
\glspl{gl}.
To produce reference images from this array,
we reconstruct them from the complete set of \gls{sa} acquisitions
using the corresponding \( \dasopmat \) operator for each insonification
and coherent compounding.
As \gls{sa} makes it possible to virtually focus both on transmit
and receive~\cite{Jensen_ULTRAS_2006},
it is often considered the gold standard,
producing images with a high resolution
while minimizing the level of \gls{sl} and \gls{ew} artifacts.

Furthermore,
in comparison to using more \enquote{exotic} ground-truth images
(\gls{eg}, \gls{ct}-like),
\gls{sa} images retain high-quality speckle patterns.
By ensuring that both subspaces contain speckle patterns arising
from sub-resolution scattering interferences,
we enforce the \gls{cnn} to preserve their information content.
We emphasize the fact that,
due to the assumptions considered in the
physical measurement process~\cref{eq:physical-operator}
and resulting backprojection operator~\cref{eq:physical-operator-adjoint}
used to define both \( \imvecapproxsubspace \) and \( \imvecgtsubspace \),
the resulting trained \gls{cnn} is not expected to correct artifacts arising
from neglected physical phenomena.
The focus is on reducing diffraction artifacts and preserving speckle
with an increased resolution.

\subsection{Convolutional Neural Network Architecture}%
\label{sec:methods:network-architecture}

\begin{figure*}[t]
    \centering
    \includegraphics{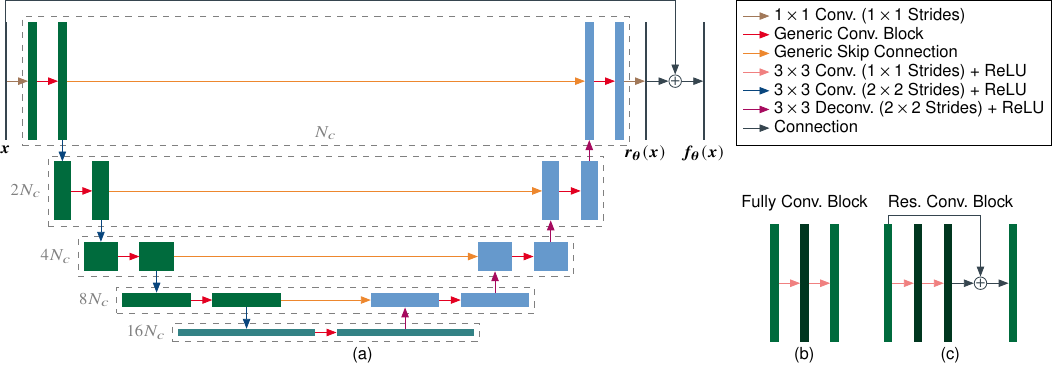}%
    {\phantomsubcaption\label{fig:methods:network-architecture:main}}%
    {\phantomsubcaption\label{fig:methods:network-architecture:fcb}}%
    {\phantomsubcaption\label{fig:methods:network-architecture:rcb}}%
    \caption{%
        Proposed residual \glsxtrfull{cnn} architecture,
        adapted from \gls{unet}~\cite{Ronneberger_MICCAI_2015}
        and from~\cite{Jin_TIP_2017} and~\cite{Perdios_IUS_2018}:
        \subref{fig:methods:network-architecture:main}
        generic overall \glsxtrshort{cnn} architecture;
        convolutional blocks considered,
        namely
        \subref{fig:methods:network-architecture:fcb}
        conventional \glsxtrfull{std-conv-block}
        and
        \subref{fig:methods:network-architecture:rcb}
        proposed \glsxtrfull{res-conv-block}.
        Connections and tensor operations (\gls{ie}, layers)
        are represented as straight,
        colorized arrows (legend in the upper right corner).
        The \ndim{3} tensors are represented as colorized rectangles.
        (Their width and height are depicted in proportion
        to the number of channels and image dimension, respectively.
        One image dimension is not represented for readability reasons.)%
    }%
    \label{fig:methods:network-architecture}
\end{figure*}

The proposed \gls{cnn} architecture
[\cref{fig:methods:network-architecture:main}]
is derived from our previous work~\cite{Perdios_IUS_2018},
adapted from the popular
\gls{unet} architecture~\cite{Ronneberger_MICCAI_2015}
and from~\cite{Jin_TIP_2017} and~\cite{Gupta_TMI_2018}.
It is a residual \gls{cnn},
expressed as \( \cnn\parens{\imvec} = \imvec + \rescnn\parens{\imvec} \),
designed to predict the negative noise to be applied to some input \( \imvec \).
It is composed of a series of multichannel \glspl{conv-layer} and \glspl{relu},
arranged in a downsampling (left arm) and upsampling (right arm) paths,
with intrinsic skip connections to mitigate information losses.
The multiscale structure confers a large receptive field to the \gls{cnn},
particularly adapted to the nonstationary restoration mapping to be learned.

The input image first undergoes a channel expansion (leftmost chamoisee arrow)
up to \( \cnnchannumb \) channels,
followed by a series of convolutional blocks (red arrows)
and downsampling layers (blue arrows) that reduce the spatial dimension
while augmenting the channel number.
The upsampling path is performed symmetrically using intrinsic skip
connections (yellow arrows),
convolutional blocks, and upsampling layers (violet arrows).
The channel number is contracted back
to its initial state (rightmost chamoisee arrow)
and the output is summed to the input image (residual skip connection).

The main differences \gls{wrt} our initial adaptation~\cite{Perdios_IUS_2018}
are as follows.
Instead of a max-pooling layer,
which seems inadequate for restoration tasks,
a \num{2 x 2} strided \gls{conv-layer} was used
within each downsampling layer as a symmetric counterpart
to the \num{2 x 2} strided \enquote{transposed} \gls{conv-layer}
used within each upsampling layer.
We used additive intrinsic skip connections instead of
concatenated ones (typical of \gls{unet} architectures),
resulting in a symmetric amount of trainable parameters in both arms.
A \gls{res-conv-block} [\cref{fig:methods:network-architecture:rcb}]
is proposed to supersede
the standard \gls{std-conv-block} [\cref{fig:methods:network-architecture:fcb}].
Note that such \glspl{res-conv-block} would not have been possible
to be deployed with concatenated skip connections directly.

\subsection{\sectitlehdrtraining}%
\label{sec:methods:hdr-training}

The trainable model parameters \( \cnnparams \) are optimized
in a supervised manner
over a training set \( \dsetpairs \) composed of \( \dsetsize \) image pairs
by minimizing the empirical risk
\begin{align}
    \emprisk\parens{\cnnparams} =
    \frac{1}{\dsetsize}
    \sum_{\dsetindex=1}^{\dsetsize}
    \loss\parens[\big]{
        \iter{\imvec}{\dsetindex},
        \cnn\parens[\big]{\iter{\imvecapprox}{\dsetindex}}
    }
    ,
    \label{eq:empirical-risk}
\end{align}
where \( \loss\parens{\imvec, \prediction{\imvec}} \)
is a nonnegative real-valued (training) loss function,
which measures the distance between a prediction
\( \prediction{\imvec} = \cnn\parens{\imvecapprox} \)
and its true value \( \imvec \).
Common loss functions include the \gls{mse},
\(
    \mseloss\parens{\imvec, \prediction{\imvec}}
    =
    \parens{1 / \imvecdim} \twonorm{\imvec - \prediction{\imvec}}^2
\),
and the \gls{mae},
\(
    \maeloss\parens{\imvec, \prediction{\imvec}}
    =
    \parens{1 / \imvecdim} \onenorm{\imvec - \prediction{\imvec}}
\).

Due to the inherent \gls{hdr} property of \gls{us} images,
they are commonly compressed (after envelope detection) before being displayed
for interpretation.
To account for the \gls{hdr} property of \gls{us} images
while preserving their \gls{rf} nature,
we introduce the \gls{mslae},
inspired by both the logarithmic compression applied to visualize \gls{us} images
and audio-coding companding algorithms
(pulse code modulation).
The associated loss is expressed as
\(
    \mslaeloss\parens{\imvec, \prediction{\imvec}}
    =
    \parens{1 / \imvecdim}
    \onenorm{ \slt\parens{\imvec} - \slt\parens{\prediction{\imvec}} }
\),
where \( \sltdef \)
is a signed (clipped-and-scaled) logarithmic transform
defined element-wise as
\begin{align}
    \sltransformeq
    \label{eq:sl-transform}
    ,
\end{align}
where \( \sltparam \in \parens{0, 1} \)
and \( \imcomponent \) is an element of \( \imvec \)
(\gls{eg}, a pixel value).
It should be noted that
\( \sltsf \parens{\imcomponent} = 0 \)
\( \forall \abs{\imcomponent} < \sltparam \).
Thus, \( \sltparam \) can be interpreted as a threshold parameter below which a
(pixel) value is assumed \enquote{negligible.}
As such,
\( \sltparam \) must be selected carefully based on the statistics
of the dataset considered
(\cref{sec:experiments:dataset:images}).
The most important feature of \gls{mslae} is that,
for any true value \( \imscl_{\imcompindex} \in \realnumbers \)
and prediction
\(
    \prediction{\imscl}_{\imcompindex}
    =
    \perturb \imscl_{\imcompindex} \in \realnumbers
\)
such that
\(
    \abs{\imscl_{\imcompindex}}, \abs{\perturb \imscl_{\imcompindex}}
    >
    \sltparam
\),
and \( \perturb > 0 \),
the resulting loss value
is a (positive) constant.
(Detailed derivations and analyses are provided in the Supplementary Material,
\cref{sec:sup:methods:hdr-training}.)
Consequently, a specific error ratio between a predicted value and its true
counterpart is penalized equally, regardless of the true value magnitude.
This is a highly desirable feature when working on \gls{hdr} data,
as is the case for (\gls{rf}) \gls{us} images.

Other log-compressed loss functions were proposed in the context of
deep learning and \gls{us}.
In~\cite{Hyun_UFFC_2019},
conventional loss functions were computed on log-compressed images
(\gls{ie}, limited to nonnegative data).
In~\cite{Luijten_TMI_2020},
the signed-mean-squared-logarithmic error (SMSLE) was introduced
as a loss function.
Because the SMSLE operates separately on the positive and negative parts
of \gls{rf} signals,
it is limited to inputs and predictions that oscillate identically
(\gls{ie}, cannot account for sign errors, making it unusable
in the present study).
Also,
both losses have a singularity at zero and can become highly unstable
as (pixel) values tend to zero.

\section{\sectitleexperiments}%
\label{sec:experiments}

\subsection{Imaging Configurations}%
\label{sec:experiments:imaging-configurations}

The imaging configurations considered in this study
(\cref{tab:imaging-configurations})
are based on the \gls{ge9ld} transducer
(\gls{ge-healthcare}, \gehealthcareaddress{})
and the \gls{vantage256} system
(\gls{verasonics}, \verasconicsaddress{})
specifications.
The \gls{ge9ld} is a \geelemnumber{}-element linear array
with a center frequency of \gecenterfreq{}
and a bandwidth of \gebandwidth{} (at \SI{-6}{\decibel}).
The transmit excitation is a single-cycle tristate waveform
of \SI{67}{\percent} duty cycle centered at \( \vsxtxfreq \),
with leading and trailing equalization pulses of quarter-cycle durations
and opposite polarities.
The received echo signals are sampled
at \( \vsxrxfreq \) (\SI{200}{\percent} bandwidth sampling).

\begin{table}[t]
    \tablefontstyle
    \centering
    \caption{Specifications of the Imaging Configurations Considered}%
    \label{tab:imaging-configurations}%
    \newcommand*{\tnoteuqphysmark}{a}
\newcommand*{\tnoteelemwidthmark}{b}
\newcommand*{\tnoteeqpulsemark}{c}
\begin{threeparttable}
\begin{tabular}{l c c c}
    \toprule
    \tableheaderstyle{Parameter}
    & \tableheaderstyle{\glsxtrshort{lq}}
    & \tableheaderstyle{\glsxtrshort{hq}}
    & \tableheaderstyle{\glsxtrshort{uq}}\tnote{\tnoteuqphysmark}
    \\
    \midrule
    Center frequency
    & \gecenterfreq{}
    & \gecenterfreq{}
    & \gecenterfreq{}
    \\
    Bandwidth
    & \gebandwidth{}
    & \gebandwidth{}
    & \gebandwidth{}
    \\
    Aperture
    & \geaperture{}
    & \geaperture{}
    & \geaperture{}
    \\
    Element number
    & \geelemnumber{}
    & \geelemnumber{}
    & \num{383}
    \\
    Pitch
    & \gepitch{}
    & \gepitch{}
    & \SI{115}{\micro\meter}\tnote{\tnoteuqphysmark}
    \\
    Element width\tnote{\tnoteelemwidthmark}
    & \geelemwidth{}
    & \geelemwidth{}
    & \geelemwidth\tnote{\tnoteuqphysmark}
    \\
    Element height
    & \geelemheight{}
    & \geelemheight{}
    & \geelemheight{}
    \\
    Elevation focus
    & \geelevationfocus{}
    & \geelevationfocus{}
    & \geelevationfocus{}
    \\
    Transmit frequency
    & \vsxtxfreq{}
    & \vsxtxfreq{}
    & \vsxtxfreq{}
    \\
    Excitation cycles\tnote{\tnoteeqpulsemark}
    & \num{1}
    & \num{1}
    & \num{1}
    \\
    Transmit-receive scheme
    & \num{1}~\glsxtrshort{pw}
    & \geelemnumber~\glsxtrshort{sa}
    & \num{383}~\glsxtrshort{sa}
    \\
    Sampling frequency
    & \vsxrxfreq{}
    & \vsxrxfreq{}
    & \vsxrxfreq{}
    \\
    \bottomrule
\end{tabular}
\begin{tablenotes}
    \item[\tnoteuqphysmark]
        \glsxtrshort{uq} is not physically possible and can only
        be simulated.
    \item[\tnoteelemwidthmark]
        Guessed (no official data available).
    \item[\tnoteeqpulsemark]
        Single excitation cycle with equalization pulses.
\end{tablenotes}
\end{threeparttable}%
\end{table}

We introduce two \enquote{natural} imaging configurations,
namely low quality (\glsunset{lq}\gls{lq})
and high quality (\glsunset{hq}\gls{hq}),
defined by the properties of the \gls{ge9ld}.
A single \gls{pw} with normal incidence and without apodization is transmitted
in the \gls{lq} configuration.
The complete set of \geelemnumber{} \gls{sa} measurements are used for \gls{hq}.
Assuming a typical speed of sound in soft tissue
of \SI{1540}{\meter\per\second},
the element spacing (\gls{ie}, pitch)
in \gls{lq} and \gls{hq} configurations is of
\( \num{\sim{} 0.78} \wavelength \)
(\gls{ie}, \( {>}\wavelength / 2 \)).
Hence,
images reconstructed by conventional \gls{das}-based algorithms will inevitably
be contaminated by \gls{gl} artifacts.
Based on the \gls{hq} configuration,
we introduce the virtual \gls{uq} one.
It takes advantage of a spatially oversampled aperture
with a halved pitch (\( \num{\sim 0.39} \wavelength \)),
resulting in a virtual \num{383}-element array,
guaranteeing \gls{gl}-free images.
To obtain the same speckle patterns as with
the \gls{hq} configuration while removing \gls{gl} artifacts,
the same aperture and geometric properties of the elements were kept.

For each imaging configuration considered,
the images were reconstructed using the corresponding
backprojection-based \gls{das} operator \( \dasopmat \)
(\cref{sec:methods:theory,sec:methods:overview})
for which the scalar weighting functions
\( \sirapproxtx \) and \( \sirapproxrx \)
and the delay functions \( \delaytx \) and \( \delayrx \)
need to be specified.
In case of \gls{pw} acquisitions (\gls{ie}, \gls{lq}),
an idealized wavefront was assumed as transmitter,
namely \( \sirapproxtx\parens{\fieldpoint} = 1 \).
In case of \gls{sa} acquisitions (\gls{ie}, \gls{hq} and \gls{uq}),
each transmission was performed with a different transducer element.
The diffraction effect of a narrow element
evaluated at a field point \( \fieldpoint \)
can be derived from a \ndim{2} far-field assumption
considering a soft baffle boundary condition
as~\cite{Delannoy_JAP_1979b,Selfridge_APL_1980}
\begin{align}
    \sirapproxtx\parens{\fieldpoint}
    =
    \frac{
        \elemwidth
        \sinc\parens{\elemwidth / \wavelength \sin\parens{\elemangle}}
    }{
        \sqrt{2 \pi} \twonorm{\fieldpoint - \sirapproxtxpos}^{1/2}
    }
    \cos\parens{\elemangle},
    \label{eq:selfridge}
\end{align}
where \( \sirapproxtxpos \) is the position of the transducer element,
\( \elemangle \) is the angle between the element normal and the vector
\( \fieldpoint - \sirapproxtxpos \),
\( \elemwidth \) is the width of the element,
and \( \sinc\parens{x} \coloneq \sin\parens{\pi x} / \parens{\pi x} \).
The transducer elements are also the receivers for all imaging configurations
and \cref{eq:selfridge} was also used
to evaluate \( \sirapproxrx\parens{\fieldpoint} \),
with \( \sirapproxrxpos \) the position of the receiving element.
The time delay functions \( \delaytx \) and \( \delayrx \)
were computed from the distance traveled by the wavefront
from the transmitter to a field point \( \fieldpoint \)
and from a field point \( \fieldpoint \) to the receiver,
respectively,
divided by the mean sound speed.

The interpolation of element raw-data values (before summation)
was performed using a \gls{bspline} approximation
of degree three~\cite{Thevenaz_TMI_2000}.
Analytic (complex) images,
often called \gls{iq} images,
were reconstructed from the analytic raw-data signals,
enabling us to have direct access to the \gls{rf} (real part)
and envelope (modulus) image representations.
The process was implemented with \gls{pyus},\footnote{\pyusurl}
a \gls{gpu}-accelerated \gls{python} package for \gls{us} imaging
developed in our laboratory.

The images were reconstructed with a width spanning the
\gls{ge9ld} aperture (\cref{tab:imaging-configurations})
and a depth from \SIrange{1}{60}{\milli\meter}.
A \( \imagesampling \) (Cartesian) grid was used to guarantee Nyquist sampling
of \gls{rf} images in both dimensions,
resulting in images of \( \numimdim \) pixels.
Examples of the resulting nonstationary \glspl{psf}
are given in the Supplementary Material
(\cref{sec:sup:experiments:imaging-configurations}).

\subsection{Simulated Dataset}%
\label{sec:experiments:dataset}

A total of \num{\datasetsize} simulated images were generated
for all imaging configurations (\cref{sec:experiments:imaging-configurations}).
Highly diverse numerical phantoms were generated
(\cref{sec:experiments:dataset:phantoms})
to produce images characterized by fully developed speckle zones
of random shapes and mean echogenicities spanning a wide dynamic range
of \SI{80}{\decibel}.
Realistic element raw-data were generated in a reasonable time frame
(\cref{sec:experiments:dataset:raw-data})
using an in-house \gls{gpu}-accelerated simulator implementing the exact
\gls{sir} model described in~\cref{eq:sir-pe-integral-model}.
Images were reconstructed using the corresponding backprojection operators
and normalized consistently (\cref{sec:experiments:dataset:images}).
The total computing time required to generate a single sample
for all imaging configurations
was of \SI{\sim1500}{\second} on a single \gls{1080ti} \gls{gpu}.
The complete dataset was simulated on multiple \glspl{gpu} for about six weeks.

\subsubsection{Simulation Phantoms}%
\label{sec:experiments:dataset:phantoms}

\begin{figure}[t]
    \centering
    \includegraphics{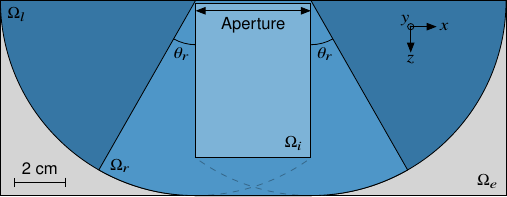}%
    \caption{%
        Representation (in the transducer plane) of the spatial domains
        used for generating the simulation phantoms for the
        imaging configurations considered (\cref{tab:imaging-configurations}).%
    }%
    \label{fig:experiments:synthetic-phantoms}
\end{figure}

Each simulation phantom is composed of a set of scatterers
defined by their positions and amplitudes.
\Cref{fig:experiments:synthetic-phantoms} shows geometric considerations
relevant to the generation of these phantoms.
The domains
\(
    \phtdomimg \subset \phtdomred \subset \phtdomlim \subset \phtdomext
    \subset \realnumbers^{3}
\)
are defined by the transducer aperture and the acquisition schemes.
All scattered contributions to the image domain \( \phtdomimg \)
arise from \( \phtdomlim \),
which is bounded by the Cartesian domain \( \phtdomext \).
To mitigate the computing time,
we only considered contributions arising from \( \phtdomred \),
where
\(
    \maxangle = \SI[round-mode=places, round-precision=2]{\maxangleval}{\degree}
\)
is the angle from which the element sensitivity falls below \SI{-6}{\decibel}.

To obtain fully developed speckle patterns throughout the resulting images,
the positions of scatterers were drawn from a uniform distribution
over \( \phtdomext \)
and their amplitudes were drawn from a normal distribution.
We computed the most restrictive (\ndim{3}) resolution cell,
defined by the \gls{fwhm} in all dimensions,
for the \gls{uq} imaging configuration,
namely \( \rescell \).
To mitigate the computing time,
an average of \sprctext{} scatterers per resolution cell was used
(\gls{ie}, lower bound in~\cite[Sec.~8.4.4]{Szabo_BOOK_2014})
and a single resolution cell was considered in elevation.
This resulted in a total of \num{\sim\scatterernumber} scatterers
in \( \phtdomred \)
(\gls{ie}, \num{\sim\sprcmmcubed} scatterers/\si{\milli\meter\cubed}).

Two hundred ellipsoidal inclusion zones were incorporated with random positions
and orientations,
with sizes of their semiaxes drawn randomly and uniformly between
\( \rescellx \) and \( \num{71} \wavelength \).
A mean echogenicity drawn randomly and uniformly between
\SIrange[
    range-phrase={ and }
]{\datasetlowermeandb}{\datasetuppermeandb}{\decibel}
\gls{wrt} the background
was set in each zone by scaling the amplitude of scatterers contained within it.

\subsubsection{Element Raw-Data Generation}%
\label{sec:experiments:dataset:raw-data}

\Cref{eq:sir-pe-integral-model} can be accurately evaluated using
the well-known \gls{fieldii} simulator~\cite{Jensen_MBEC_1996, Jensen_UFFC_1992},
but computing time requirements were prohibitive.
To enable the generation of a sufficiently large dataset,
we implemented an in-house \glsxtrshort{gpu}-accelerated
simulator~\cite{Perdios_ARXIV_2021}.
The main differences compared with \gls{fieldii}
are the spline-based representations used for both time and element-surface
domains.
\Gls{nurbs} representations and Gauss-Legendre quadrature are used
for surface integrations,
enabling high accuracy with few integration points.
The time domain is represented in a \gls{bspline} basis~\cite{Thevenaz_TMI_2000},
reducing the sampling frequency requirements.

The transducer elements were represented by cylindrical
\gls{nurbs} surfaces,
and \( \num{3 x 87} \) quadrature points were used for the surface integral.
Their electromechanical impulse response was approximated by a differentiated
log-normal-windowed sine wave.
A soft boundary condition was considered and a constant speed of sound of
\SI{\meansoundspeedval}{\meter\per\second} was set.
To minimize the simulation time,
we used a \gls{bspline} of degree five for the time-domain representation,
enabling sufficient accuracy (\gls{ie}, \SI{> 60}{\decibel})
with a sampling frequency of \simrxfreq{}.
The implementation has been validated against \gls{fieldii}
and enabled an overall \num{200}-fold speed-up.

\subsubsection{Normalization and Statistical Considerations}%
\label{sec:experiments:dataset:images}

For each imaging configuration,
a normalization factor was determined on
independent realizations of a reference simulation phantom composed of
scatterers resulting in a constant mean echogenicity of \SI{0}{\decibel}.
Assuming a fully developed speckle zone of constant mean echogenicity,
image envelope values follow a Rayleigh distribution~\cite{Wagner_TSU_1983}.
Thus,
for a Rayleigh distributed speckle with a \num{0}-\si{\decibel} mean echogenicity,
an interval of \SIrange{-12}{+6}{\decibel}
covering \SI{\datasetvalinterval}{\percent}
of the envelope values can be determined
(Supplementary Material,
\cref{sec:sup:methods:rayleigh-statistics:confidence-interval}).
Since the simulation phantoms contain inclusions of constant mean echogenicities
ranging from \SIrange{\datasetlowermeandb}{\datasetuppermeandb}{\decibel},
an interval of \SIrange{\datasetlowervaldb}{\datasetuppervaldb}{\decibel}
(\gls{ie}, \num{\datasetrangevaldb}-\si{\decibel} range)
was considered.

\begin{figure*}[t]
    \centering
    \includegraphics{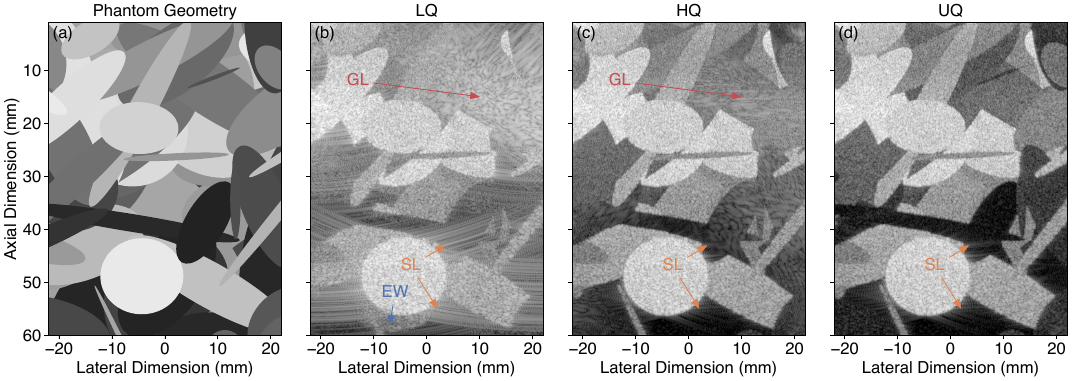}%
    {\phantomsubcaption\label{fig:experiments:sim-pht-sample:mask}}%
    {\phantomsubcaption\label{fig:experiments:sim-pht-sample:lq}}%
    {\phantomsubcaption\label{fig:experiments:sim-pht-sample:hq}}%
    {\phantomsubcaption\label{fig:experiments:sim-pht-sample:uq}}%
    \caption{%
        \Gls{bmode} image representations
        (\num{\datasetrangevaldb}-\si{\decibel} range)
        of a simulated dataset sample:
        \subref{fig:experiments:sim-pht-sample:mask}
        the phantom mask composed of elliptical inclusions;
        images reconstructed using each imaging configuration considered
        (\cref{tab:imaging-configurations}),
        namely
        \subref{fig:experiments:sim-pht-sample:lq}
        \glsxtrfull{lq} configuration,
        \subref{fig:experiments:sim-pht-sample:hq}
        \glsxtrfull{hq} configuration
        (\gls{ie}, gold-standard image for the physical transducer array),
        and
        \subref{fig:experiments:sim-pht-sample:uq}
        \glsxtrfull{uq} configuration
        (\gls{ie}, gold-standard image for the spatially oversampled virtual
        version of the transducer array, considered as ground-truth).
        Some zones dominated by \glsxtrfull{gl},
        \glsxtrfull{sl},
        and \glsxtrfull{ew} artifacts
        are highlighted by colorized arrows and associated annotations.%
    }%
    \label{fig:experiments:sim-pht-sample}
\end{figure*}

Representations of a simulated dataset sample are shown
in \cref{fig:experiments:sim-pht-sample},
highlighting the differences in diffraction artifacts between
\gls{lq}, \gls{hq}, and \gls{uq} configurations
[\cref{fig:experiments:sim-pht-sample:lq,fig:experiments:sim-pht-sample:hq,%
fig:experiments:sim-pht-sample:uq}].
The image obtained from the \gls{uq} configuration
is free from \gls{gl} artifacts,
while the \gls{hq} image still suffers from them
(although significantly reduced compared with \gls{lq}).
Both \gls{hq} and \gls{uq} configurations result in images free from
\gls{ew} artifacts and with \gls{sl} artifacts significantly reduced
compared with \gls{lq}.
Due to the spatial dependency of the imaging configuration \glspl{psf}
(Supplementary Material, \cref{sec:sup:experiments:imaging-configurations}),
the spread of these artifacts is also spatially dependent.

\subsection{Training and Hyperparameter Search}%
\label{sec:experiments:training-setup}

Many training experiments were performed for hyperparameter search
(Supplementary Material, \cref{sec:sup:hyperparameters}).
For each training experiment,
kernel weights were initialized using the well-known \gls{glorot}
initialization~\cite{Glorot_AISTATS_2010} with a uniform distribution,
and biases were initialized to zero.
Model parameters were optimized using the Adam optimizer~\cite{Kingma_ARXIV_2014}
with a learning rate of \num{5e-5}.
Mini-batch learning was deployed with a batch size of two.
The training set consisted of \num{\trainsetsize} image pairs;
a size motivated by a dedicated study to prevent training experiments
from overfitting
(Supplementary Material, \cref{sec:sup:hyperparameters:training-size}).
A total of \num{\iterationnumber} iterations were performed,
corresponding to \num{\sim{} 33} epochs.
Complete random shuffling of the training set was performed
after each epoch.
Neither training regularization
(\gls{eg}, dropout or weight regularization)
nor data augmentation was used.
To fulfill the downsampling restrictions imposed by the
proposed \gls{cnn} architecture (\cref{fig:methods:network-architecture}),
input images were zero padded symmetrically to the closest supported
image shape (\gls{ie}, \( \numimdimpad \)),
and cropped to their original size after inference.

To monitor and evaluate the performance of each training experiment,
we used a validation set of \num{\validsetsize} image pairs.
Both the \gls{psnr} and the \gls{ssim} index~\cite{Wang_TIP_2004}
were computed at each validation step
(\gls{ie}, every \num{\stepsperepoch} iterations).
These metrics were evaluated on \gls{bmode} representations between
\SIrange[
    range-phrase={ and }
]{\datasetlowervaldb}{\datasetuppervaldb}{\decibel}
(confidence interval discussed in \cref{sec:experiments:dataset:images}),
and were averaged over the entire validation set.
The \gls{bmode} \gls{ssim} correlated well with visual assessments
for evaluating the overall quality of recovered images,
and was used to select the best performing \gls{cnn} instance among
the \num{500} validation steps of each training experiment.
For comparison purposes,
a fixed random seed was used for initializing kernel weights,
identical training set shufflings were performed,
and the same validation set was used for each training experiment.

The implementation was carried out using
\gls{tensorflow}\footnote{\tfurl} (\tfver),
and the trainings were performed on \gls{v100} \glspl{gpu}.

\subsection{\exptitlenumerical}%
\label{sec:experiments:numerical-evaluations}

\Glsxtrlong{us} image quality is conventionally assessed using metrics
reflecting lesion detectability,
such as the contrast,
the \gls{cnr},
or the resolution of the imaging system~\cite{Liebgott_IUS_2016}.
As demonstrated in~\cite{Rindal_UFFC_2019},
\glspl{dra},
which are common to most adaptive beamformers,
may improve contrast or \gls{cnr} measures
without actually improving lesion detectability,
and may even conceal information relevant to clinical diagnosis.
Inspired by~\cite{Rindal_UFFC_2019},
we designed a dedicated numerical test phantom
composed of tissue-mimicking echogenic zones
embedded in an anechoic background
[\cref{fig:results:numerical-phantom:mask}].
Each zone is described hereafter
in conjunction with the associated metrics.

A block with a square section of
\SI{20 x 20}{\milli\meter}
is centered at (\SI{-5}{\milli\meter}, \SI{20}{\milli\meter}).
A low-echogenic cylindrical inclusion with a diameter of \SI{8.5}{\milli\meter}
is embedded at its center.
The contrast between the two is of \SI{-36}{\decibel}
such that the diffraction artifacts covering the low-echogenic inclusion
of \gls{lq} images [\cref{fig:results:numerical-phantom:lq}]
are significantly higher (\SI{\sim{} 8}{\decibel}) than the inclusion level.
The restoration quality of the low-echogenic inclusion was assessed
by computing the contrast~\cite{Smith_UMB_1985},
expressed in decibels as
\(
    \text{C}
    =
    20 \logten\parens{
        \expectedvalue\bracks{\vec{s}_{\metricdominclsymb}}
        /
        \expectedvalue\bracks{\vec{s}_{\metricdomblocksymb}}
    },
\)
where \( \vec{s}_{\metricdomblocksymb} \) and \( \vec{s}_{\metricdominclsymb} \)
are the envelope-detected image amplitude values in \( \metricdomblock \)
and \( \metricdomincl \),
respectively,
and \( \expectedvalue\bracks{\argdot} \) is the expected value,
evaluated as the sample mean
(Supplementary Material,
\cref{sec:sup:methods:rayleigh-statistics:contrast}).
By considering an inclusion with a prescribed contrast,
reconstruction errors (\gls{eg}, \glspl{dra})
would most likely result in erroneous contrast estimates.

Another block with a rectangular section of
\( \geaperture \times \SI{10}{\milli\meter} \)
(\gls{ie}, spanning the probe aperture)
is positioned at a depth of \SI{50}{\milli\meter},
characterized by a lateral log-linear echogenicity ranging
from \SIrange{+30}{-50}{\decibel}
(\gls{ie}, \num{\sim{} 1.82} \si{\decibel\per\milli\meter}).
The capacity of the proposed method to preserve prescribed linearity
(while removing artifacts)
was assessed by averaging the obtained image
amplitudes within \( \metricdomlg \) along the depth axis,
and the accuracy was visually assessed by comparing it with the prescribed one.
Potential \glspl{dra} would result in (highly) distorted
amplitude gradients.

Four ideal bright reflectors
(\( \metricreflectorA \), \( \metricreflectorB \), \( \metricreflectorC \),
and \( \metricreflectorD \))
are arranged at a lateral position of
\SI{12.5}{\milli\meter} and depths of \SIlist{10;20;30;40}{\milli\meter}.
Both axial and lateral \gls{fwhm} measures were evaluated
on the image amplitude using a \ndim{2} spline-based interpolation
and a sub-pixel peak finder
within \( \num{2}\wavelength \times \num{2}\wavelength \)
regions centered at the position of each bright reflector.

Speckle patterns were assessed within a square region (\( \metricdomsr \))
of size \( \num{10}\wavelength \times \num{10}\wavelength \),
centered at (\SI{0}{\milli\meter}, \SI{27}{\milli\meter}).
First-order statistics was evaluated by computing the ratio between the mean
and the standard deviation of image amplitudes,
often referred to as \gls{snr},
expressed as
\(
    \text{\glsxtrshort{snr}}
    =
    \expectedvalue\bracks{\vec{s}_{\metricdomsrsymb}}
    /
    \parens{\variance\parens{\vec{s}_{\metricdomsrsymb}}}^{1/2}
\).
In the case of samples following a Rayleigh distribution
(\gls{ie}, fully developed speckle),
this ratio would be equal to \num{1.91}~\cite{Burckhardt_TSU_1978}
(Supplementary Material,
\cref{sec:sup:methods:rayleigh-statistics:first-order}).
Second-order statistics was evaluated by
computing the \gls{fwhm} of the \ndim{2}
\gls{acf}~\cite{Foster_UI_1983,Wagner_TSU_1983}
(Supplementary Material,
\cref{sec:sup:methods:rayleigh-statistics:second-order}).
This metric represents a statistical measure of the
\enquote{speckle resolution,}
in both axial and lateral dimensions,
and is of great importance to many post-processings
(\gls{eg}, speckle tracking).

The level of diffraction artifacts was quantified by averaging the
image amplitudes within different anechoic rectangular regions.
These regions were selected on \gls{lq} images
[\cref{fig:results:numerical-phantom:lq}]
to be dominated by significant diffraction artifacts primarily caused by
\glspl{gl} (\( \metricdomgl \)),
\glspl{sl} (\( \metricdomsl \)),
and \glspl{ew} (\( \metricdomew \)).
Global \gls{psnr} and \gls{ssim} metrics were also computed
on \gls{bmode} images between
\SIrange[
    range-phrase={ and }
]{\datasetlowervaldb}{\datasetuppervaldb}{\decibel}
against \gls{uq} images.

Three hundred statistically independent realizations
(\gls{ie}, random scatterers)
were generated identically to the simulated dataset
(\cref{sec:experiments:dataset}).
An additional normalization factor was evaluated on the average
of all \gls{uq} test images such that the reconstructed gradient would fit
(on average) the prescribed one.
This factor was applied to all images of each imaging configurations
(\gls{ie}, also before inference).
No renormalization was applied after inference.

From the hyperparameter search carried out
(Supplementary Material, \cref{sec:sup:hyperparameters}),
four trained \glspl{cnn} were selected for evaluations
using the numerical test phantom.
To evaluate the effect of the training loss function,
we considered three instances of the proposed residual \gls{cnn},
deployed with \glspl{res-conv-block},
additive intrinsic skip connections,
and \num{16} initial expansion channels
(\cref{fig:methods:network-architecture}),
and trained using \gls{mse} (\resnumcnnAlgd),
\gls{mae} (\resnumcnnBlgd),
and \gls{mslae} (\resnumcnnClgd)
as loss functions.
A \num{32}-channel instance,
trained using \gls{mslae} as loss function (\resnumcnnDlgd),
was also selected to evaluate the effect of increasing network capacity.
The proposed \gls{mslae},
defined in \cref{eq:sl-transform},
was implemented with a \enquote{threshold} parameter \( \sltparam \)
corresponding to \SI{\datasetlowervaldb}{\decibel}
(confidence interval, \cref{sec:experiments:dataset:images}).

\subsection{\exptitleexperimental}%
\label{sec:experiments:experimental-evaluations}

Experimental data were acquired using a \gls{ge9ld} transducer
on a \gls{vantage256} system using the imaging configurations
defined in \cref{sec:experiments:imaging-configurations}
(\cref{tab:imaging-configurations}),
except for the \gls{uq} case
(simulation exclusive).
Compounded acquisitions were performed at maximum \gls{prf}
(\gls{ie}, \SI{\sim{} 9.5}{\kilo\hertz})
to minimize the effect of potential inter-acquisition motion.
The single \gls{pw} insonification (\gls{lq}) was performed first
in the ultrafast sequence,
directly followed by \geelemnumber{} \gls{sa} acquisitions (\gls{hq}),
performed in an alternated manner from central to outer elements.
A peak-to-peak voltage of \SI{50}{\volt} was used for the transmit excitation.
\Gls{tgc} was implemented to compensate for a mean tissue attenuation
of \SI{-0.5}{\decibel\per\centi\meter\per\mega\hertz}.

\Gls{invitro} acquisitions were carried out on a \gls{cirs-054gs}
general-purpose ultrasound phantom (\gls{cirs}, \cirsaddress{}).
The transducer was clamped on a stand during acquisitions
and its face was immersed in water for acoustic coupling.
A normalization factor was determined
in the same manner as described in \cref{sec:experiments:dataset:images},
for both \gls{lq} and \gls{hq} imaging configurations
on fully developed speckle zones of the \gls{invitro} phantom.
These normalization factors were applied to all images reconstructed from
experimental acquisitions (including before inference).
Quantitative metrics were evaluated on a zone of the phantom composed of
three circular inclusions with a radius of \SI{4}{\milli\meter}
and centered at a depth of \SI{40}{\milli\meter}
[\cref{fig:results:experimental:pht-hypo-lq,%
fig:results:experimental:pht-hypo-cnn,%
fig:results:experimental:pht-hypo-hq}]:
an anechoic inclusion (\( \metricexpdomcia \))
and two low-echogenic inclusions of
\SI{-6}{\decibel} (\( \metricexpdomcib \))
and \SI{-3}{\decibel} (\( \metricexpdomcic \)).
For each inclusion,
the contrast was computed against a background zone (\( \metricexpdombckg \)).
Speckle patterns were assessed using first- and second-order statistics
(\cref{sec:experiments:numerical-evaluations})
within a square region (\( \metricexpdomsr \)) of
\( \num{10}\wavelength \times \num{10}\wavelength \)
centered at (\SI{0}{\milli\meter}, \SI{27}{\milli\meter}).

An \gls{invivo} sequence of \num{60} frames was acquired
at a frame rate of \SI{30}{\hertz}
on the carotid of a volunteer.
The transducer was positioned on the neck of the volunteer to image
a longitudinal view of the right carotid.
Acoustic coupling was achieved by applying a layer of conventional
\gls{us} coupling gel.
All images within the \gls{invivo} sequence were reconstructed identically,
with the normalization factors
evaluated on the \gls{invitro} phantom.

Experimental acquisitions were evaluated on images obtained with
the proposed approach using the trained \resexpcnnAlgd{} \gls{cnn},
and compared with \gls{lq} images (\gls{cnn} inputs)
and \gls{hq} images (considered as references).
We opted for a \gls{cnn} deployed with \num{16} initial expansion channels
because of its real-time inference capabilities
(Supplementary Material, \cref{tab:results:trainings:inference-timings}).

\section{Results}%
\label{sec:results}

\subsection{\exptitlenumerical}%
\label{sec:results:numerical-evaluations}

\begin{figure*}[t]
    \centering
    \includegraphics{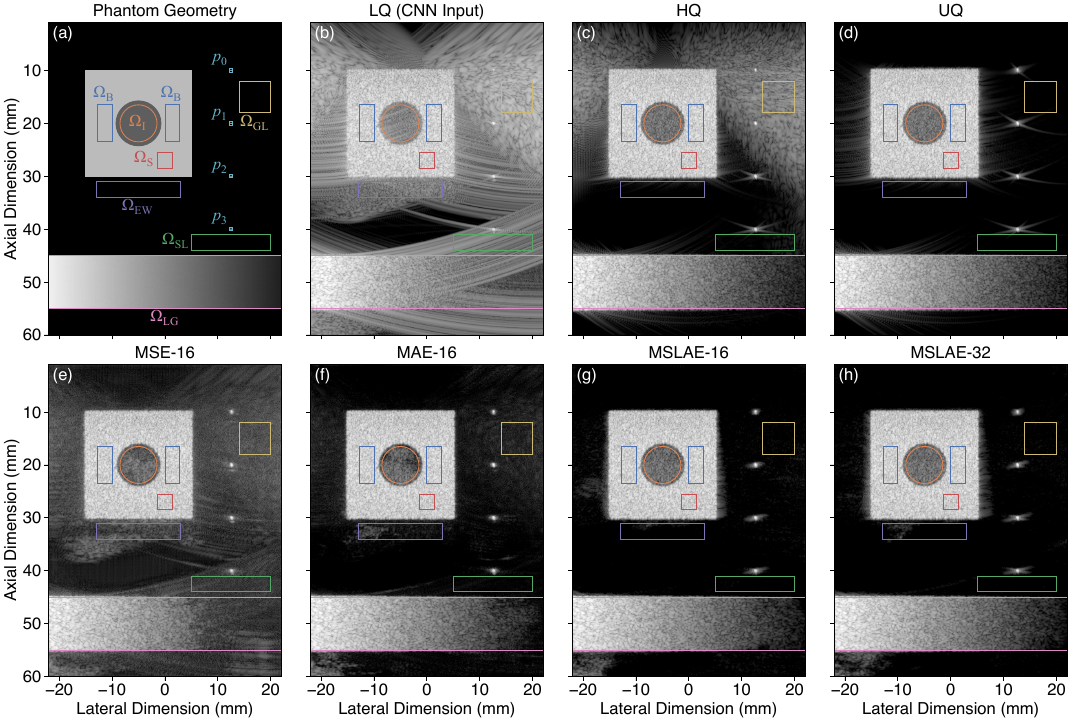}%
    {\phantomsubcaption\label{fig:results:numerical-phantom:mask}}%
    {\phantomsubcaption\label{fig:results:numerical-phantom:lq}}%
    {\phantomsubcaption\label{fig:results:numerical-phantom:hq}}%
    {\phantomsubcaption\label{fig:results:numerical-phantom:uq}}%
    {\phantomsubcaption\label{fig:results:numerical-phantom:cnnA}}%
    {\phantomsubcaption\label{fig:results:numerical-phantom:cnnB}}%
    {\phantomsubcaption\label{fig:results:numerical-phantom:cnnC}}%
    {\phantomsubcaption\label{fig:results:numerical-phantom:cnnD}}%
    \caption{%
        \Gls{bmode} image representations
        (\num{\datasetrangevaldb}-\si{\decibel} range)
        of a numerical test phantom sample:
        \subref{fig:results:numerical-phantom:mask}
        the phantom mask and annotated zones
        in which the metrics were evaluated;
        images reconstructed using each imaging configuration considered
        (\cref{tab:imaging-configurations}),
        namely
        \subref{fig:results:numerical-phantom:lq}
        \glsxtrfull{lq} configuration,
        \subref{fig:results:numerical-phantom:hq}
        \glsxtrfull{hq} configuration
        (\gls{ie}, gold-standard image for the physical transducer array),
        and
        \subref{fig:results:numerical-phantom:uq}
        \glsxtrfull{uq} configuration
        (\gls{ie}, reference image);
        images recovered from the \glsxtrshort{lq} input image
        using the proposed approach
        with each of the trained \glsxtrfullpl{cnn} considered
        (\cref{sec:experiments:numerical-evaluations}),
        namely
        \subref{fig:results:numerical-phantom:cnnA}
        \resnumcnnAlgd{},
        \subref{fig:results:numerical-phantom:cnnB}
        \resnumcnnBlgd{},
        \subref{fig:results:numerical-phantom:cnnC}
        \resnumcnnClgd{},
        and
        \subref{fig:results:numerical-phantom:cnnD}
        \resnumcnnDlgd{}.%
    }%
    \label{fig:results:numerical-phantom}
\end{figure*}

\begin{table*}[t]
    \tablefontstyle
    \centering
    \caption{Numerical Test Phantom Metrics}%
    \label{tab:numerical-phantom-metrics}%
    \newcommand*{\tnotemetricsreport}{a}
\newcommand*{\tnotecontrastmetric}{b}
\newcommand*{\tnotepsnrssim}{c}

\newcommand{\tablemetricmainskip}{\enskip\enskip\enskip}
\newcommand{\tablemetricinterskip}{~}
\begin{threeparttable}
\begin{tabular}{
    ll
    r
    @{\tablemetricinterskip}
    l
    r
    @{\tablemetricinterskip}
    l
    r
    @{\tablemetricinterskip}
    l
    r
    @{\tablemetricinterskip}
    l
    r
    @{\tablemetricinterskip}
    l
    r
    @{\tablemetricinterskip}
    l
    r
    @{\tablemetricinterskip}
    l
}
    \toprule
    \multicolumn{2}{l}{\tableheaderstyle{Metric}\tnote{\tnotemetricsreport}}
    & \multicolumn{2}{c}{\tableheaderstyle{\glsxtrshort{lq}}}
    & \multicolumn{2}{c}{\tableheaderstyle{\glsxtrshort{hq}}}
    & \multicolumn{2}{c}{\tableheaderstyle{\glsxtrshort{uq}}}
    & \multicolumn{2}{c}{\tableheaderstyle{\resnumcnnAlgd}}
    & \multicolumn{2}{c}{\tableheaderstyle{\resnumcnnBlgd}}
    & \multicolumn{2}{c}{\tableheaderstyle{\resnumcnnClgd}}
    & \multicolumn{2}{c}{\tableheaderstyle{\resnumcnnDlgd}}
    \\
    \midrule
    \multicolumn{2}{l}{C (\si{\decibel})\tnote{\tnotecontrastmetric}}
    & \num{-28.33} & (\num{0.60})
    & \num{-36.06} & (\num{0.32})
    & \num{-36.06} & (\num{0.32})
    & \num{-39.41} & (\num{0.83})
    & \num{-39.49} & (\num{1.03})
    & \num{-37.74} & (\num{0.70})
    & \num{-37.40} & (\num{0.51})
    \\
    \multicolumn{2}{l}{\glsxtrshort{gl} (\si{\decibel})}
    & \num{+6.39} & (\num{0.45})
    & \num{-10.27} & (\num{0.52})
    & \num{-66.62} & (\num{0.46})
    & \num{-27.99} & (\num{0.32})
    & \num{-45.34} & (\num{0.35})
    & \num{-61.56} & (\num{0.36})
    & \num{-62.24} & (\num{0.32})
    \\
    \multicolumn{2}{l}{\glsxtrshort{sl} (\si{\decibel})}
    & \num{-14.48} & (\num{1.08})
    & \num{-58.16} & (\num{0.55})
    & \num{-68.64} & (\num{0.38})
    & \num{-38.89} & (\num{0.54})
    & \num{-60.49} & (\num{0.53})
    & \num{-67.51} & (\num{0.52})
    & \num{-67.91} & (\num{0.64})
    \\
    \multicolumn{2}{l}{\glsxtrshort{ew} (\si{\decibel})}
    & \num{-13.64} & (\num{0.30})
    & \num{-59.61} & (\num{0.67})
    & \num{-59.96} & (\num{0.70})
    & \num{-31.23} & (\num{0.78})
    & \num{-46.74} & (\num{3.04})
    & \num{-60.74} & (\num{4.05})
    & \num{-55.17} & (\num{6.55})
    \\
    \midrule
    \multicolumn{2}{l}{\glsxtrshort{snr}}
    & \num{1.84} & (\num{0.09})
    & \num{1.80} & (\num{0.09})
    & \num{1.80} & (\num{0.09})
    & \num{1.80} & (\num{0.09})
    & \num{1.80} & (\num{0.09})
    & \num{1.79} & (\num{0.09})
    & \num{1.81} & (\num{0.09})
    \\
    \multicolumn{2}{l}{\glsxtrshort{acf} lat.\@ (\si{\micro\meter})}
    & \num{262.1} & (\num{19.5})
    & \num{219.4} & (\num{15.2})
    & \num{219.6} & (\num{15.3})
    & \num{245.6} & (\num{17.5})
    & \num{246.2} & (\num{17.6})
    & \num{251.1} & (\num{18.4})
    & \num{246.0} & (\num{17.4})
    \\
    \multicolumn{2}{l}{\glsxtrshort{acf} ax.\@ (\si{\micro\meter})}
    & \num{293.5} & (\num{21.5})
    & \num{302.6} & (\num{21.4})
    & \num{302.7} & (\num{21.4})
    & \num{301.5} & (\num{21.6})
    & \num{302.9} & (\num{21.9})
    & \num{301.9} & (\num{21.6})
    & \num{301.8} & (\num{21.6})
    \\
    \midrule
    \multirow{4}{*}{\rotatebox[origin=c]{90}{\glsxtrshort{fwhm} lat.\@}}
    & \( \metricreflectorA \) (\si{\micro\meter})
    & \num{276.6} & (\num{24.4})
    & \num{202.1} & (\num{1.8})
    & \num{202.2} & (\num{0.0})
    & \num{226.5} & (\num{8.5})
    & \num{211.5} & (\num{6.9})
    & \num{232.5} & (\num{7.9})
    & \num{207.3} & (\num{4.1})
    \\
    & \( \metricreflectorB \) (\si{\micro\meter})
    & \num{336.2} & (\num{5.6})
    & \num{242.5} & (\num{0.9})
    & \num{242.7} & (\num{0.0})
    & \num{255.2} & (\num{3.2})
    & \num{243.1} & (\num{2.6})
    & \num{270.1} & (\num{6.3})
    & \num{240.5} & (\num{1.9})
    \\
    & \( \metricreflectorC \) (\si{\micro\meter})
    & \num{388.6} & (\num{1.5})
    & \num{280.0} & (\num{0.0})
    & \num{280.5} & (\num{0.0})
    & \num{286.3} & (\num{2.0})
    & \num{293.7} & (\num{1.6})
    & \num{301.1} & (\num{1.6})
    & \num{271.3} & (\num{0.9})
    \\
    & \( \metricreflectorD \) (\si{\micro\meter})
    & \num{446.6} & (\num{4.9})
    & \num{321.9} & (\num{0.0})
    & \num{322.4} & (\num{0.0})
    & \num{345.9} & (\num{6.1})
    & \num{340.5} & (\num{4.9})
    & \num{359.4} & (\num{2.2})
    & \num{322.5} & (\num{2.2})
    \\
    \midrule
    \multirow{4}{*}{\rotatebox[origin=c]{90}{\glsxtrshort{fwhm} ax.\@}}
    & \( \metricreflectorA \) (\si{\micro\meter})
    & \num{264.8} & (\num{8.3})
    & \num{266.5} & (\num{0.8})
    & \num{266.6} & (\num{0.0})
    & \num{265.0} & (\num{2.9})
    & \num{256.3} & (\num{2.3})
    & \num{286.2} & (\num{5.7})
    & \num{241.6} & (\num{2.4})
    \\
    & \( \metricreflectorB \) (\si{\micro\meter})
    & \num{316.7} & (\num{2.5})
    & \num{314.6} & (\num{0.3})
    & \num{314.6} & (\num{0.0})
    & \num{313.0} & (\num{2.4})
    & \num{310.8} & (\num{1.6})
    & \num{308.5} & (\num{3.3})
    & \num{312.3} & (\num{1.5})
    \\
    & \( \metricreflectorC \) (\si{\micro\meter})
    & \num{317.5} & (\num{0.9})
    & \num{318.3} & (\num{0.0})
    & \num{318.3} & (\num{0.0})
    & \num{314.7} & (\num{0.6})
    & \num{311.7} & (\num{0.5})
    & \num{303.5} & (\num{1.3})
    & \num{311.5} & (\num{0.6})
    \\
    & \( \metricreflectorD \) (\si{\micro\meter})
    & \num{320.7} & (\num{1.6})
    & \num{324.0} & (\num{0.0})
    & \num{324.0} & (\num{0.0})
    & \num{324.8} & (\num{2.2})
    & \num{328.4} & (\num{1.5})
    & \num{319.3} & (\num{1.6})
    & \num{312.7} & (\num{1.1})
    \\
    \midrule
    \multicolumn{2}{l}{\glsxtrshort{psnr} (\si{\decibel})}
    & \num{8.60} & (\num{0.04})
    & \num{14.23} & (\num{0.04})
    & \multicolumn{2}{c}{\invalidtabentry\tnote{\tnotepsnrssim}}
    & \num{14.71} & (\num{0.05})
    & \num{21.96} & (\num{0.12})
    & \num{24.18} & (\num{0.25})
    & \num{25.14} & (\num{0.32})
    \\
    \multicolumn{2}{l}{\glsxtrshort{ssim}}
    & \num{0.31} & (\num{0.00})
    & \num{0.73} & (\num{0.00})
    & \multicolumn{2}{c}{\invalidtabentry\tnote{\tnotepsnrssim}}
    & \num{0.39} & (\num{0.00})
    & \num{0.58} & (\num{0.00})
    & \num{0.75} & (\num{0.00})
    & \num{0.78} & (\num{0.00})
    \\
    \bottomrule
\end{tabular}
\begin{tablenotes}
    \item[\tnotemetricsreport]
    Metrics were averaged over \num{\numtestsetsize} independent realizations.
    The standard deviation is given in parentheses.
    \item[\tnotecontrastmetric]
    Prescribed contrast of \SI{-36}{\decibel}.
    \item[\tnotepsnrssim] \glsxtrshort{psnr} and \glsxtrshort{ssim} metrics
    were computed against \glsxtrshort{uq}.
\end{tablenotes}
\end{threeparttable}%
\end{table*}

Visual assessment of the test phantom images
(\cref{fig:results:numerical-phantom})
and the metrics obtained
(\cref{tab:numerical-phantom-metrics})
confirm that the proposed image reconstruction method
significantly improves the image quality compared with \gls{lq}
using any of the trained \glspl{cnn}.
A global comparison of \glspl{cnn} with identical capacities
(\gls{ie}, \resnumcnnAlgd{}, \resnumcnnBlgd{}, and \resnumcnnClgd{})
demonstrates the superiority of the proposed \gls{hdr}-sensitive
\gls{mslae} loss.
Both \gls{mslae} trainings achieved higher global quality metrics
(\gls{ie}, \gls{psnr} and \gls{ssim})
than \gls{hq},
mainly due to the impressive reduction of \gls{gl} artifacts.
The added capacity of \resnumcnnDlgd{} resulted
in improved overall performances compared with \resnumcnnClgd{}.

The restoration of the low-echogenic inclusion
(prescribed contrast of \SI{-36}{\decibel})
and the resulting contrast obtained were improved drastically
compared with \gls{lq}.
Both \resnumcnnAlgd{} and \resnumcnnBlgd{}
suffer from important
\enquote{dark region artifacts}~\cite{Rindal_IUS_2017}
in the low-echogenic inclusion
[\( \metricdomincl \) in
\cref{fig:results:numerical-phantom:cnnB,fig:results:numerical-phantom:cnnC}],
whereas \resnumcnnClgd{} and \resnumcnnDlgd{} provide a more accurate
restoration of the inclusion.
This is confirmed by the contrast obtained
which are tending to the reference one (\gls{ie}, \gls{uq})
for \resnumcnnClgd{} and \resnumcnnDlgd{}
(\cref{tab:numerical-phantom-metrics}).

All trained \glspl{cnn} resulted in diffraction artifact levels
drastically reduced compared with \gls{lq}.
Remaining \gls{gl} artifacts were far below \gls{hq} (\SI{> 18}{\decibel}).
Artifacts caused by \glspl{ew} appeared to be the most complex
artifact to deal with
[\gls{eg}, bottom-left corner in
\cref{fig:results:numerical-phantom:cnnA,%
fig:results:numerical-phantom:cnnB,%
fig:results:numerical-phantom:cnnC,%
fig:results:numerical-phantom:cnnD}].
It can also be observed that the restoration of the \gls{sl} artifacts present
in the \gls{uq} reference image was more accurate with
\resnumcnnDlgd{}.

The \gls{snr} obtained within the speckle zone \( \metricdomsr \)
for \gls{das}-based methods (\gls{ie}, \gls{lq}, \gls{hq}, and \gls{uq})
did not reach the theoretical value of \num{1.91} for fully developed speckle.
This was expected since \sprctext{} scatterers per resolution cell
were used for numerical simulations
(\gls{ie}, lower bound to obtain fully developed speckle).
All trained \glspl{cnn} improved the \gls{snr} compared with \gls{lq}
(\gls{ie}, closer to the \gls{uq} one).
The lateral resolution of speckle patterns
(\cref{tab:numerical-phantom-metrics}, \gls{acf} lat.)
was only slightly improved,
without reaching the one of \gls{hq} and \gls{uq}.
On the other hand,
the lateral resolution evaluated on bright reflectors
was improved significantly.

\begin{figure}[t]
    \centering
    \includegraphics{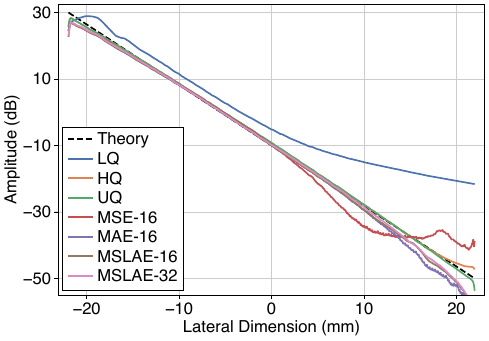}%
    \caption{%
        Mean amplitude responses (averaged along the axial dimension)
        of the horizontal gradient zone in the numerical phantom
        [\( \metricdomlg \) in \cref{fig:results:numerical-phantom:mask}],
        averaged over \num{\numtestsetsize} independent realizations.%
    }%
    \label{fig:results:numerical-phantom-gradient}
\end{figure}

The restoration results of the log-linear gradient are shown in
\cref{fig:results:numerical-phantom-gradient}.
Almost perfect restoration was achieved from \SIrange{+30}{-30}{\decibel},
with a slight but increasing deviation for lower echogenicity values,
by all trained \glspl{cnn} except for \resnumcnnAlgd{},
which only preserved linearity from \SIrange{+30}{-15}{\decibel}.
Note that the \gls{lq} response is slightly overestimated (offset)
because of the ideal \gls{pw} assumption used
to derive the backprojection (\gls{das}) operator
(\cref{sec:experiments:imaging-configurations}).
It is easily resolved by all trained \glspl{cnn}.

An additional representation obtained from the incoherent averaging
(\gls{ie}, after envelope detection)
of all test images is provided in the Supplementary Material
(\cref{sec:sup:results:numerical-evaluations}),
with an emphasis on remaining artifacts.

\subsection{\exptitleexperimental}%
\label{sec:results:experimental-evaluations}

\begin{figure*}[t]
    \centering
    \includegraphics{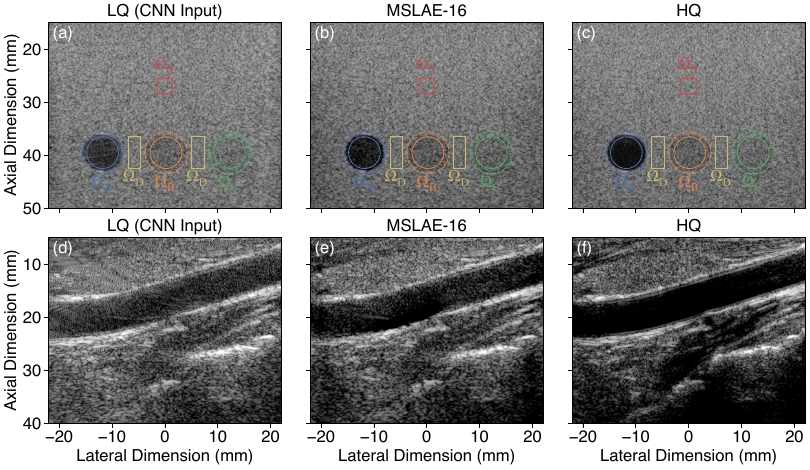}%
    {\phantomsubcaption\label{fig:results:experimental:pht-hypo-lq}}%
    {\phantomsubcaption\label{fig:results:experimental:pht-hypo-cnn}}%
    {\phantomsubcaption\label{fig:results:experimental:pht-hypo-hq}}%
    {\phantomsubcaption\label{fig:results:experimental:caro-lq}}%
    {\phantomsubcaption\label{fig:results:experimental:caro-cnn}}%
    {\phantomsubcaption\label{fig:results:experimental:caro-hq}}%
    \caption{%
        \Gls{bmode} image representations
        of an \gls{invitro} test phantom
        (top row, \num{78}-\si{\decibel} range)
        and an \gls{invivo} carotid sample
        (bottom row, \num{50}-\si{\decibel} range):
        \subref{fig:results:experimental:pht-hypo-lq}
        and
        \subref{fig:results:experimental:caro-lq}
        single \glsxtrfull{pw} \glsxtrfull{lq} images;
        \subref{fig:results:experimental:pht-hypo-cnn}
        and
        \subref{fig:results:experimental:caro-cnn}
        images recovered from \glsxtrshort{lq} using the proposed
        \glsxtrfull{cnn}-based image reconstruction method
        with the selected trained \glsxtrshort{cnn}
        (\gls{ie}, \resnumcnnClgd{});
        \subref{fig:results:experimental:pht-hypo-hq}
        and
        \subref{fig:results:experimental:caro-hq}
        reference \glsxtrfull{hq} images reconstructed from the complete set
        of \glsxtrfull{sa} acquisitions.%
    }%
    \label{fig:results:experimental}
\end{figure*}

\Cref{fig:results:experimental} shows the experimental results of
an example image for both \gls{invitro} (top row)
and \gls{invivo} (bottom row) acquisitions.
Overall,
it can be observed that,
despite using only simulated data for training,
the key effects of the proposed approach translated well to experimental settings.

The visual assessment of \gls{invitro} results
shows that both \gls{sl} and \gls{gl} artifacts
(clearly visible in the anechoic inclusion)
were strongly reduced.
(Note that \gls{ew} artifacts are harder to identify as they result
in patterns that resemble speckle.)
This effect was confirmed quantitatively by the contrast measured in
each inclusion of the \gls{invitro} phantom
(\cref{tab:experimental-phantom-metrics}).
The contrast in the anechoic inclusion was largely improved compared with
\gls{lq}.
However, the proposed approach seemed to slightly \enquote{overshoot}
in the other two inclusions compared with \gls{hq} (reference).
Speckle patterns were generally well-preserved.
Yet,
almost no improvement in the lateral resolution of speckle patterns
was observed and measured (\cref{tab:experimental-phantom-metrics}).

\begin{table}[t]
    \tablefontstyle
    \centering
    \caption{Experimental Test Phantom Metrics}%
    \label{tab:experimental-phantom-metrics}%
    \newcommand*{\tnotemetricsreport}{a}
\newcommand*{\tnotepsnrssim}{b}

\newcolumntype{M}[1]{>{\raggedright\let\newline\\\arraybackslash\hspace{0pt}}m{#1}}
\newcolumntype{N}[1]{>{\centering\let\newline\\\arraybackslash\hspace{0pt}}m{#1}}
\newcolumntype{O}[1]{>{\raggedleft\let\newline\\\arraybackslash\hspace{0pt}}m{#1}}

\newcommand{\tablemetricmainskip}{\enskip\enskip\enskip}
\newcommand{\tablemetricinterskip}{~}
\begin{tabular}{
    l
    >{\centering\arraybackslash}m{1.5cm}
    >{\centering\arraybackslash}m{1.5cm}
    >{\centering\arraybackslash}m{1.5cm}
}
    \toprule
    \multicolumn{1}{l}{\tableheaderstyle{Metric}}
    & \multicolumn{1}{c}{\tableheaderstyle{\glsxtrshort{lq}}}
    & \multicolumn{1}{c}{\tableheaderstyle{\glsxtrshort{hq}}}
    & \multicolumn{1}{c}{\tableheaderstyle{\resnumcnnClgd}}
    \\
    \midrule
    \multicolumn{1}{l}{\( \text{C}_{\text{A}} \) (\si{\decibel})}
    & \num{-19.77}
    & \num{-29.55}
    & \num{-25.93}
    \\
    \multicolumn{1}{l}{\( \text{C}_{\text{B}} \) (\si{\decibel})}
    & \num{-6.18}
    & \num{-6.30}
    & \num{-6.71}
    \\
    \multicolumn{1}{l}{\( \text{C}_{\text{C}} \) (\si{\decibel})}
    & \num{-2.77}
    & \num{-3.41}
    & \num{-4.04}
    \\
    \midrule
    \multicolumn{1}{l}{\glsxtrshort{snr}}
    & \num{1.93}
    & \num{1.91}
    & \num{1.92}
    \\
    \multicolumn{1}{l}{\glsxtrshort{acf} lat.\@ (\si{\micro\meter})}
    & \num{282.3}
    & \num{220.6}
    & \num{281.0}
    \\
    \multicolumn{1}{l}{\glsxtrshort{acf} ax.\@ (\si{\micro\meter})}
    & \num{284.2}
    & \num{279.7}
    & \num{291.0}
    \\
    \bottomrule
\end{tabular}%
\end{table}

The \gls{invivo} experiments cover the full complexity of \gls{us} imaging,
namely highly diverse scattering processes,
a wide range of echogenicities,
and all physical effects neglected in the simulated dataset used for training.
Yet,
diffraction artifacts were strongly reduced,
especially visible in zones where \gls{sl} and \gls{gl} artifacts aggregate
[\gls{eg}, top left of \cref{fig:results:experimental:caro-lq}].
Structures initially shadowed by such artifacts were well restored,
up to some degree of residual artifacts.
Image quality improvements were less visible in deeper regions,
partially due to the fact that diffraction artifacts do not seem
to be dominant there.
Very fine and low-echogenic details, such as the carotid intima,
were not accurately restored.
An overall remaining clutter noise was observed,
for instance within the carotid or in the anechoic regions below it
(also in the \gls{lq} case).
We computed standard image quality metrics,
namely the \gls{psnr}, the \gls{ssim}, and the contrast between the tissue and the lumen.
However, both the \gls{psnr} and the \gls{ssim} are strongly image-dependent
and therefore not suitable for comparing different experiments.
Also, contrast measures between \glspl{roi} in the tissue and in the lumen
vary greatly depending on the choice of these \glspl{roi}
and the level of artifacts contained within.
We therefore chose not to report them
and limited ourselves to a qualitative analysis.
The complete \gls{invivo} sequence is presented in video format
(Supplementary Material).

\section{Discussion}%
\label{sec:discussion}

\subsection{Performance in Ideal Conditions}%
\label{sec:discussion:ideal-settings}

The potential of the proposed \gls{cnn}-based image reconstruction method
was demonstrated through the results obtained in numerical experiments
(\cref{sec:results:numerical-evaluations}),
in which the physical assumptions of the \gls{sir} model,
defined in \cref{eq:sir-pe-integral-model}
and used to simulate the training dataset,
were fully satisfied.
These results showed that the proposed method
is capable of strongly reducing (nonstationary) diffraction artifacts,
mainly caused by \glspl{gl}, \glspl{sl}, and \glspl{ew},
while preserving speckle patterns that result from main lobes.
Moreover,
it is capable of accurately recovering zones initially hidden by
diffraction artifacts,
on a dynamic range exceeding \SI{60}{\decibel}.
This means that the detectability of lesions potentially hidden
by such artifacts would be largely improved.
The use of simulated reference images obtained from an optimal version
of the linear transducer considered (\gls{ie}, \gls{uq}),
together with the \gls{hdr}-sensitive and \gls{rf}-compatible
\gls{mslae} training loss,
enabled the reconstruction of images from single-\gls{pw} acquisitions
with a quality similar to that of (gold-standard) \gls{sa} imaging.
This represents a more than 100-fold reduction in acquisition requirements,
such as acquisition time,
power consumption,
or data transfer rates.

Artifacts caused by \glspl{ew} were the most difficult to tackle,
most likely due to their close resemblance to speckle patterns.
This issue may be addressed using a \gls{cnn} with greater capacity.
Also, \gls{ew} artifacts could be reduced
by a suitable transmit apodization~\cite{Udesen_UFFC_2008,JensenJonas_UFFC_2016},
at the cost of a lower insonification energy on the sides of the field of view
(probably restorable if accounted for in the training set),
and an increased transmitter complexity.
The lateral resolution of speckle patterns was only slightly improved
compared with \gls{lq}.
This complex task was better achieved with \resnumcnnDlgd{} than \resnumcnnClgd{},
suggesting that a greater capacity \gls{cnn} could improve further
the tightening of speckle patterns.

Many elements composing the numerical test phantom
were not present in the training dataset.
Training samples were formed by random ellipses of constant mean echogenicity
filling out the entire image domain.
Hence,
no rectilinear boundaries,
isolated bright reflectors,
anechoic zones,
or echogenicity gradients
were seen during training.
The trained \glspl{cnn} were robust to these (unseen) features,
suggesting that the complex restoration mapping involved in the proposed method
was learned accurately.
In particular,
the robustness observed on bright reflectors strongly suggests that
the learned mapping is not limited to fully developed speckle zones
that composed the simulated-image dataset.

It was also observed that when using a \gls{cnn} with increased capacity
(\gls{ie}, \resnumcnnClgd{} \gls{vs} \resnumcnnDlgd{}),
not only the metrics were improved,
but also the restoration of remaining diffraction artifacts present
in the \gls{uq} reference images (\gls{ie}, \glspl{sl}).
This confirms that the learning of the restoration mapping
(from \gls{lq} to \gls{uq})
is effective,
and may be achieved exactly with a \gls{cnn} of even greater capacity,
provided that a sufficiently large training dataset is available
to avoid overfitting.

\subsection{Performance in Experimental Conditions}%
\label{sec:discussion:experimental-settings}

\Gls{invitro} experiments showed that images were improved significantly
over conventional single \gls{pw} (\gls{lq}) images.
A reduction of diffraction artifacts was also observed
on \gls{invivo} acquisitions,
in particular at shallow depths.
Yet,
performance drops were observed compared with numerical evaluations
performed in ideal conditions.
Such performance drops were expected because the \glspl{cnn}
were trained exclusively on simulated data.
They are likely to be caused by all physical phenomena
not accurately represented in the training dataset.

A first set of potential differences come from the transducer array itself.
Indeed, not all parameters were known or possible to be measured accurately.
In particular,
the electromechanical impulse response and exact geometry of each piezoelectric
forming the transducer array could only be approximated.
These parameters have an influence on the resulting (spatially dependent)
\gls{psf} of the imaging system.

Another important set of differences come from the physical assumptions
inherent to the \gls{sir} model considered for both simulating the dataset
and deriving the backprojection operator
(\cref{sec:methods:theory,sec:methods:overview}).
The \gls{sir} model only accounts for diffusive scattering
in the medium,
and hence neither specular nor diffractive scattering regimes were taken into
account~\cite[Sec.~8.2]{Szabo_BOOK_2014}.
Such scattering regimes result in image statistics deviating
from purely diffusive (Rayleigh) ones~\cite{Wagner_TSU_1983,Tuthill_UI_1988}
and may therefore disrupt trained models not accounting for such
statistical features.
Speed of sound is also assumed constant
in the \gls{sir} model,
and hence deviations in mean speed of sound and/or local fluctuations
may alter speckle patterns (\gls{ie}, image statistics).
Dispersive attenuation was compensated using a standard \gls{tgc},
which only corrects for a constant and frequency-independent attenuation.
This may represent a limitation,
especially at great depths,
because the frequency-dependent attenuation continuously distorts
the acoustic pulse as it travels through the medium.
Thermal noise and quantization noise also have an increased impact with depth,
as the backscattered signal amplitude decreases.
While less likely to have a significant impact
in ultrafast acquisitions (low mechanical index),
non-linear effects could also result in discrepancies.

Although \ndim{3} simulations were performed,
only extruded \ndim{2} phantoms were considered on a layer of \( \rescelly \),
namely approximately one-sixth of the transducer height.
This choice was made for computational reasons,
but it means that potential out-of-plane artifacts
were not exactly accounted for in the training dataset.

\subsection{Potential Improvements}%
\label{sec:discussion:limitations}

As the training dataset is a crucial component of the method,
it also represents a great area for possible improvements.
In ideal test conditions
(\cref{sec:discussion:ideal-settings}),
we observed that
the deployed \gls{cnn} trained using the proposed \gls{mslae}
as loss function is even capable of partially restoring
remaining artifacts of \gls{uq} reference images (\gls{ie}, \glspl{sl}).
Using reference images with a quality even higher than \gls{uq}
(\gls{eg}, ideal \glspl{psf}) could be considered.
As the main reason for performance drops in experimental conditions
seems to be related to the physical phenomena neglected for simulating
the training dataset,
the use of more sophisticated simulations and/or experimental training datasets
could lead to improved results in experimental conditions.
Experimental datasets are of interest
as acquisition time is reduced compared to simulation
and all physical phenomena are taken into account.
Yet,
undesirable physical phenomena (\gls{eg}, frequency-dependent attenuation)
also impact reference images,
and the acquisition of such an experimental dataset
with high diversity and free of motion artifacts
is a challenging task.
More sophisticated simulated datasets are also appealing.
They could contain (low-quality) input images suffering from
diffraction artifacts and other undesirable physical phenomena,
and (high-quality) reference images free of these.

The proposed \gls{cnn}-based image reconstruction method
(\cref{sec:methods:overview})
relies on a backprojection operator.
This operator
is a \gls{das} algorithm with weighting (apodization) functions that result from
the (far-field) physical assumptions made.
While deviating from the theoretical derivations,
the proposed method could also be implemented with common apodization functions
(\gls{eg}, a Hamming window),
conventionally designed to reduce diffraction artifacts
at the cost of a lower lateral image resolution.
One should keep in mind that the trained \glspl{cnn}
were more efficient at reducing diffraction artifacts
than improving the lateral resolution of speckle patterns.
Also,
the backprojection-based \gls{das} operator results in
a \gls{psf} with a tighter main lobe and higher diffraction artifacts
than more \enquote{restrictive} apodization functions commonly used.

The fact that the restoration mapping
is learned on a specific imaging configuration
(\gls{ie}, array geometry, impulse response, transmit wavefront, \gls{etc})
theoretically limits its use to said configuration.
While this limits the approach,
it maximizes its potential as the entire \gls{cnn} capacity
is used to learn an already complex nonstationary restoration problem.
Also, it is common in \gls{us} imaging systems to
have finely tuned image reconstruction settings for each imaging configuration.
Datasets accounting for variations in some imaging configuration parameters
could be considered,
and would probably enable a greater generalization while degrading performances.

As opposed to regularization techniques,
the proposed approach does not contain an explicit data fidelity
feedback mechanism.
Data fidelity is \enquote{only} inferred implicitly by the (supervised)
training strategy.
Combinations of optimization algorithms and learned projections
could be considered (\gls{eg}, \cite{Gupta_TMI_2018}),
at the risk of losing real-time imaging capabilities.

Among all hyperparameter searches carried out
(supplementary material, \cref{sec:sup:hyperparameters})
the use of the proposed \gls{hdr}-sensitive and \gls{rf}-compatible
\gls{mslae} as loss function provided the largest increase in performance.
Other architectural and/or optimization parameters could be optimized.
An in-depth study of the activation function
would be of particular interest in the context of (oscillating) \gls{rf} signals
because of the asymmetric (positive) nature of \gls{relu}.
Preliminary studies conducted on this aspect using anti-rectifier-like
activations did not yield satisfactory results so far.

\subsection{Application Perspectives}%
\label{sec:discussion:perspectives}

The proposed approach may provide a viable solution
to ultrafast \gls{us} imaging modes
(\gls{eg}, shear-wave elastography)
in settings where only a few acquisitions are possible to track the underlying
(fast-evolving) physical phenomena accurately,
and where diffraction artifacts can severely degrade the accuracy
of these imaging modes~\cite{Montaldo_UFFC_2009}.
It should be noted that such ultrafast imaging modes
rely on the time-coherence of moving speckle patterns
between consecutive frames,
and that static image metrics used in this work cannot assess such a coherence.
A preliminary quantitative study was carried out on the latter aspect
with positive outcomes~\cite{Perdios_IUS_2019}.
Visual assessment of the \gls{invivo} carotid sequence
(Supplementary Material, video)
also suggests that the time-coherence of moving speckle patterns is preserved.
We recently demonstrated that the proposed method
enables accurate displacement estimations in zones initially shadowed
by diffraction artifacts,
in both numerical and \gls{invivo} conditions~\cite{Perdios_TMI_2021}.

Portable systems could also benefit from the proposed approach
to reduce the number of transmit-receive events required per frame,
and reach more efficient power-down states of some
electronic components~\cite{Hager_UFFC_2019}.
Also,
the complexity of the transmitter could be reduced
as beamforming is unnecessary to transmit unfocused wavefronts.
Sparse-array imaging could also benefit from the proposed approach
as subsampling the transducer aperture inevitably entails substantial
increases in diffraction artifacts.
Preliminary results in the context of single-\gls{pw} imaging using sparse
linear arrays were recently presented~\cite{Perdios_IUS_2020c}.

It is also interesting to note that
the use of a backprojection-based operation
(akin to \gls{das} in its computational complexity)
followed by an inference
is readily compatible with real-time imaging
(Supplementary Material, \cref{tab:results:trainings:inference-timings}).

\section{Conclusion}%
\label{sec:conclusion}

We proposed a \gls{cnn}-based image reconstruction method
for high-quality ultrafast \gls{us} imaging.
A low-quality estimate is obtained by means of a backprojection operation,
akin to conventional \gls{das} beamforming,
from which a high-quality image is then restored using a \gls{cnn}
trained specifically to remove diffraction artifacts.
Trainings were performed on a simulated dataset using
a loss function designed to account for both the \gls{hdr} and the oscillating
properties of \gls{rf} \gls{us} images.
Through extensive numerical experiments,
we demonstrated that the proposed method can effectively reconstruct images
from single \gls{pw} insonifications with a quality comparable
to that of \gls{sa} imaging.
This represents a more than 100-fold reduction in acquisition requirements,
which could unlock ultrafast imaging modalities
where only single insonifications are possible,
or could enable a significant reduction in power consumption of portable systems.
\Gls{invitro} and \gls{invivo} experiments confirmed that trainings carried out
on simulated images perform well in experimental settings.
Yet,
dedicated datasets could improve performances in experimental settings
or enable the learning of even more complex restoration mappings.
The proposed method is readily compatible with real-time imaging.
It could also benefit to other acquisition and imaging systems relying
on antenna arrays and suffering from diffraction artifacts.

\section*{Acknowledgment}%
\label{sec:acknowledgment}

The authors would like to thank warmly the many contributors to this work:
Adrien Besson
for the initial discussions and contributions,
Olivier Bernard
for the discussions on \gls{us}-specific image metrics and speckle quality,
Arthur Père and Jérémie Gringet for their invaluable
help in implementing and benchmarking hundreds of models,
and for the many suggestions for improvements,
Malo Grisard and Philippe Rossinelli for their help in the early stages
of this work,
\Gls{verasonics} for the technical support on the \gls{vantage256} system,
and
the editors and anonymous reviewers for their helpful comments and suggestions.

\bibliographystyle{IEEEtran}
\bibliography{resources/references}

\supplements{}
\section{\sectitlemethods}%
\label{sec:sup:methods}

\subsection{Notes on Regularized Regression Techniques}
\label{sec:sup:methods:regularization}

As briefly summarized in \cref{sec:methods:theory},
regularized regression techniques may be used
to solve inverse problems of the form
\( \rdvec = \physopmat \imvec + \nvec \),
where \( \physopmatdef \) is the measurement (matrix) operator,
\( \imvecdef \) is the (vectorized) image we seek to recover,
\( \rdvecdef \) are the (vectorized) measurements
(element raw data).
They can typically be deployed to improve the image quality
as an alternative to using multiple insonifications,
and imply finding a solution
\begin{align}
    \estimate{\imvec}
    =
    \argmin_{\imvecdef}
    \braces*{
        \datafidelity\parens{\physopmat \imvec, \rdvec}
        + \regparam \regularizer\parens{\imvec}
    }
    ,
    \label{eq:ip-opt}
\end{align}
where \( \datafidelity \colon \rdvecdom \times \rdvecdom \to \realnumbers_{+} \)
is a data fidelity term (\gls{eg}, the \( \ell_{2} \)-norm),
and \( \regularizer \colon \imvecdom \to \realnumbers_{+} \)
is a regularizer used to infer prior knowledge on the expected image.
The parameter \( \regparam \in \realnumbers_{+} \)
controls the weighting of the regularization
and is typically adjusted manually.
One solution to~\cref{eq:ip-opt},
when using the \( \ell_{2} \)-norm as data fidelity,
may be found using the well-known
proximal gradient descent iteration~\cite{Combettes_BOOK_2011}
\begin{align}
    \iter{\imvec}{\iterindex + 1}
    =
    \proxregexpl\parens[\big]{
        \iter{\imvec}{\iterindex}
        - \gradientstep \physopmat\adj \physopmat \iter{\imvec}{\iterindex}
        + \gradientstep \physopmat\adj \rdvec
    }
    ,
    \label{eq:prox-gradient-descent-iter}
\end{align}
where
\( \physopmat\adj \) is the adjoint of \( \physopmat \),
\( \gradientstep \in \realnumbers_{+} \)
is the gradient step size (that may also be iteration-dependent),
and the proximity operator
\( \proxregexpl \colon \imvecdom \to \imvecdom \)
is defined as
\begin{align}
    \prox_{\proxregparam \regularizer}\parens{\vec{z}}
    =
    \argmin_{\vec{z} \in \imvecdom}
    \frac{1}{2}
    \twonorm{\imvec - \vec{z}}^2 + \proxregparam \regularizer\parens{\imvec}
    ,
    \label{eq:prox-operator}
\end{align}
for some parameter \( \proxregparam \in \realnumbers_{+} \).
The proximity operator in \cref{eq:prox-gradient-descent-iter}
acts as a projection to \enquote{denoise} each estimate based on some prior
knowledge of \( \imvec \).

One can note that if~\cref{eq:prox-gradient-descent-iter}
is initialized to zero, namely \( \imvec^{\parens{0}} = \vec{0} \),
the first estimate is obtained as
\(
    \iter{\imvec}{1}
    =
    \proxregexpl\parens{\gradientstep \physopmat\adj \rdvec}
\),
which corresponds to a backprojection followed by some \enquote{denoising}
projection that depends on the regularizer \( \regularizer \).
This observation is the basis of the proposed
\enquote{two-step} approach described in~\cref{sec:methods:overview}.

\subsection{Statistical Considerations of Fully Developed Speckle}%
\label{sec:sup:methods:rayleigh-statistics}

Ultrasound speckle is characteristic of images produced by conventional
\gls{das}-based pulse-echo imaging systems;
it arises from the coherent interferences of echo-components reflected by
sub-resolved diffusive scatterers.
It is said to be fully developed when scatterers are present in sufficient
numbers within resolution cells,
and resulting backscattered signals
follow a (circular symmetric) complex normal distribution
\( \complexnormaldist(0, 2 \sigma^{2}) \),
where \( \sigma^{2} \) represents the variance
of each (independent) component~\cite{Wagner_TSU_1983}.
Following envelope detection,
the signal amplitude of these interferences
follow a Rayleigh distribution~\cite{Burckhardt_TSU_1978},
denoted as \( \rayleighdist\parens{\sigma} \),
with a parameter \( \sigma > 0 \) related to the underlying
(circular symmetric) complex distribution.
The corresponding \gls{pdf} and \gls{cdf} are defined as
\begin{align}
    \rayleighpdf\parens{x; \sigma}
    &=
    \frac{x}{\sigma^{2}} e^{-x^{2} / \parens{2 \sigma^{2}}}
    \label{eq:rayleigh-pdf}
    ,
    \\
    \rayleighcdf\parens{x; \sigma}
    &=
    1 - e^{-x^{2} / \parens{2 \sigma^{2}}}
    \label{eq:rayleigh-cdf}
    ,
\end{align}
for \( x \geq 0 \), respectively.
The first moment (\gls{ie}, mean or expected value),
the second moment,
and the variance
of a Rayleigh random variable \( \rndvaramplitude \) are given by
\begin{align}
    \mu_{1}
    &=
    \expectedvalue\bracks{\rndvaramplitude}
    =
    \sqrt{\frac{\pi}{2}} \sigma
    \label{eq:rayleigh-first-moment}
    ,
    \\
    \mu_{2}
    &=
    \expectedvalue\bracks{\rndvaramplitude^{2}}
    =
    2 \sigma^{2}
    \label{eq:rayleigh-second-moment}
    ,
    \\
    \variance\parens{\rndvaramplitude}
    &=
    \expectedvalue\bracks{\rndvaramplitude^{2}}
    - \expectedvalue\bracks{\rndvaramplitude}^{2}
    =
    \frac{4 - \pi}{2} \sigma^{2}
    \label{eq:rayleigh-variance}
    .
\end{align}
Even though speckle patterns are sometimes interpreted as noise,
they contain positional information about the underlying physical phenomenon,
as they result from deterministic interferences,
and are therefore extensively exploited in motion analysis%
~\cite{Montaldo_UFFC_2009}.

\subsubsection{First-Order Statistics}%
\label{sec:sup:methods:rayleigh-statistics:first-order}

The analysis of first-order statistics provides useful tools to characterize
envelope signals regardless of the acquisition system geometry and are therefore
extensively used in image quality metrics and tissue characterization.
A widely used measure of first-order statistics in \gls{us} imaging is
the ratio of mean to standard deviation of a signal
(\gls{ie}, the reciprocal of the coefficient of variation),
often referred to as \gls{snr}~\cite{Burckhardt_TSU_1978}.
In the case of a signal following a Rayleigh distribution,
it is given by
\begin{align}
    \text{\glsxtrshort{snr}}
    =
    \frac{\expectedvalue\bracks{\rndvaramplitude}}{\sqrt{\variance\parens{\rndvaramplitude}}}
    =
    \sqrt{\frac{\pi}{4 - \pi}}
    \approx
    \num{1.91}
    \label{eq:rayleigh-snr}
    .
\end{align}
This ratio is ideally estimated at a single point in an image by conducting
multiple independent realizations and estimating
\( \expectedvalue\bracks{\rndvaramplitude} \) and \( \variance\parens{\rndvaramplitude} \)
using the sample mean and sample variance.
Assuming a zone of a physical domain composed of a large amount
of random diffuse scatterers with constant mean amplitude
imaged with a system
characterized by a slowly varying \gls{psf} within such a zone,
the resulting speckle patterns will inherit
quasi-constant statistical properties.
Therefore,
one can assume a \gls{wss} process within that image zone and estimate
the \gls{snr} directly from the samples obtained.
It should be noted that,
in an attempt to reduce speckle \enquote{noise,}
one may want to improve the \gls{snr} defined in \cref{eq:rayleigh-snr}
by reducing \( \variance\parens{\rndvaramplitude} \).
Yet,
in scenarios where accurate speckle patterns are required
(\gls{eg}, motion estimation),
the goal is to preserve such patterns
or to restore them as they may have been altered
by imaging artifacts or thermal noise.
The \gls{snr} can thus serve to verify that speckle patterns
follow the expected (first order) statistics.

\subsubsection{Second-Order Statistics}%
\label{sec:sup:methods:rayleigh-statistics:second-order}
To study the spatial characteristics of speckle patterns,
which depend on the \gls{psf} of the imaging system,
the evaluation of second-order statistics is required.
The normalized \gls{acf},
also referred to as \gls{pcc},
is commonly used for this purpose~\cite{Wagner_TSU_1983,Foster_UI_1983}.
Assuming a \gls{wss} process,
it is defined as
\begin{align}
    \rho_{\rndvaramplitude \rndvaramplitude}
    \parens{\Delta \impos}
    =
    \frac{
        \expectedvalue\bracks{
            \parens{
                \rndvaramplitude\parens{\impos_{1}}
                -
                \expectedvalue\bracks{\rndvaramplitude}
            }
            \parens{
                \rndvaramplitude\parens{\impos_{2}}
                -
                \expectedvalue\bracks{\rndvaramplitude}
            }\conj
        }
    }{
        \variance\parens{\rndvaramplitude}
    }
    \label{eq:normalized-acf-wss}
    ,
\end{align}
where
\( \impos_{1} \) and \( \impos_{2} \) are two positions (in the image),
\( \Delta \impos = \impos_{2} - \impos_{1} \),
and \( z\conj \) denotes the complex conjugate of \( z \).
Note that the numerator of \cref{eq:normalized-acf-wss}
is simply the autocorrelation of the image amplitudes
from which the mean was subtracted,
and may therefore be efficiently estimated using two \glspl{fft}.
To characterize the \enquote{resolution} (coarseness) of speckle patterns,
the \gls{fwhm} of the \gls{acf} is typically evaluated
in all image dimensions~\cite{Wagner_TSU_1983}.

\subsubsection{Contrast}%
\label{sec:sup:methods:rayleigh-statistics:contrast}

A commonly used definition of contrast between two \gls{us} signals
(or image zones)
\( \rndvaramplitude_{1} \) and \( \rndvaramplitude_{2} \)
is given by their ratio of mean amplitude,
expressed in decibels as~\cite{Smith_UMB_1985}
\begin{align}
    \text{C}
    =
    20
    \logten\parens[\bigg]{
        \frac{
            \expectedvalue\bracks{\rndvaramplitude_{1}}
        }{
            \expectedvalue\bracks{\rndvaramplitude_{2}}
        }
    }
    \label{eq:contrast-db-amplitude}
    .
\end{align}
Considering a Rayleigh random variable \( \rndvaramplitude \),
one can note that,
from \cref{eq:rayleigh-first-moment,eq:rayleigh-second-moment},
\(
    \expectedvalue\bracks{\rndvaramplitude}
    =
    (\pi \expectedvalue\bracks{\rndvaramplitude^{2}} / 4)^{1 / 2}
\).
Hence,
if
\( \rndvaramplitude_{1} \sim \rayleighdist\parens{\sigma_{1}} \)
and
\( \rndvaramplitude_{2} \sim \rayleighdist\parens{\sigma_{2}} \),
\cref{eq:contrast-db-amplitude} can be equivalently expressed,
in decibels,
on the signal intensity (power) as
\begin{align}
    \text{C}
    =
    10
    \logten\parens[\bigg]{
        \frac{
            \expectedvalue\bracks{\rndvaramplitude_{1}^{2}}
        }{
            \expectedvalue\bracks{\rndvaramplitude_{2}^{2}}
        }
    }
    \label{eq:contrast-db-intensity}
    .
\end{align}

When considering fully developed speckle signals
that follow a Rayleigh distribution,
the analysis of the signal intensity,
which follows an exponential distribution,
may be of interest as it is linearly proportional
to the concentration of scatterers~\cite{Smith_UMB_1985}.
However,
in general,
the signal amplitude is the quantity of interest in \gls{us} imaging
as \gls{us} systems sense \gls{rf} signals that are linearly proportional
to the amplitude of scatterers~\cite{Wagner_TSU_1983},
irrespectively of their statistical properties.
Hence,
\cref{eq:contrast-db-intensity} should not be used on signals deviating
from Rayleigh statistics as it could result in unrealistic values.

\subsubsection{Confidence Interval}%
\label{sec:sup:methods:rayleigh-statistics:confidence-interval}

\begin{figure}[t]
    \centering
    \includegraphics{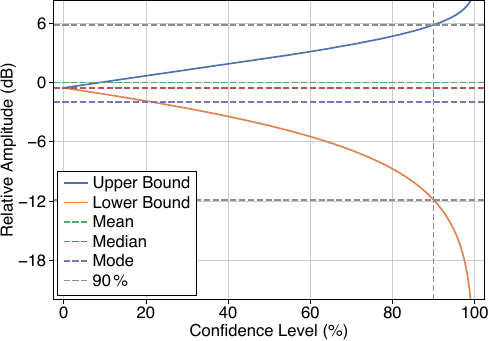}%
    \caption{
        Lower and upper confidence bounds (in decibels)
        \gls{wrt} the confidence level (in percent)
        of a Rayleigh random variable normalized by its expected value
        (\gls{ie}, mean).%
    }%
    \label{fig:methods:rayleigh-confidence-interval}
\end{figure}

As the \gls{cdf} of a Rayleigh random variable,
defined in \cref{eq:rayleigh-cdf},
is continuous and strictly monotonically increasing,
its quantile function \( \rayleighquantile = \rayleighcdf^{-1} \),
and can be expressed as
\begin{align}
    \rayleighquantile\parens{y; \sigma}
    =
    \sigma \sqrt{-2 \ln(1 - y)}
    \label{eq:rayleigh-quantile}
    ,
\end{align}
for \( y \in [0, 1) \).
Considering a symmetric confidence level \( \confidencelevel \in [0, 1) \),
the lower and upper confidence bounds are obtained directly
from~\cref{eq:rayleigh-quantile},
and expressed as
\begin{align}
    \parens*{
        \sigma \sqrt{-2 \ln\parens*{\frac{1 + \confidencelevel}{2}}}
        ,\,
        \sigma \sqrt{-2 \ln\parens*{\frac{1 - \confidencelevel}{2}}}
    }
    \label{eq:rayleigh-confidence-bounds}
    .
\end{align}
\Cref{fig:methods:rayleigh-confidence-interval} shows the confidence bounds
of a Rayleigh random variable normalized by its expected value,
namely
\(
    \rndvaramplitude / \mu_{1}
    \sim
    \rayleighdist\parens{\sqrt{2 / \pi}}
\).
A \SI{\datasetvalinterval}{\percent} confidence level is therefore achieved
when accounting values ranging from approximately
\SIrange{-12}{+6}{\decibel}
\gls{wrt} its expected value (\gls{ie}, mean).

\subsection{\sectitlehdrtraining}%
\label{sec:sup:methods:hdr-training}

Recall (from \cref{sec:methods:hdr-training})
that the proposed \gls{mslae} loss is expressed as
\begin{align}
    \mslaelosseq
    \label{eq:mslae}
    ,
\end{align}
where \( \sltdef \)
is a signed (clipped-and-scaled) logarithmic transform
defined element-wise (pixel-wise) in \cref{eq:sl-transform} as
\begin{align}
    \sltransformeq
    \label{eq:sup:sl-transform}
    ,
\end{align}
where \( \sltparam \in \parens{0, 1} \)
and \( \imcomponent \) is an element of \( \imvec \)
(\gls{eg}, a pixel value).

To anticipate the effect of the \gls{mslae} loss,
let us define
a predicted value \( \prediction{\imscl} = \perturb \imscl \)
for any true value \( \imscl \in \set{R} \)
and error ratio \( \perturb \in \set{R} \).
(Note that the component index \( \imcompindex \) has been dropped to lighten
notation.)
The resulting loss function can be expressed as
\begin{align}
    &
    \mslaeloss\parens{\imscl , \perturb\imscl}
    =
    \abs{\slt \parens{\imscl} - \slt\parens{\perturb\imscl}}
    \nonumber
    \\
    &
    \quad
    =
    \begin{cases}
        \abs*{ \log_{\sltparam}\parens{\perturb} }
        &
        \text{for}\:
        \abs{\imscl} > \sltparam,\:
        \abs{\perturb\imscl} > \sltparam,\:
        \perturb > 0
        ,
        \\
        \abs*{
            \log_{\sltparam}\parens{
                -\sltparam^{2} / \parens{\perturb \imscl^{2}}
            }
        }
        &
        \text{for}\:
        \abs{\imscl} > \sltparam,\:
        \abs{\perturb\imscl} > \sltparam,\:
        \perturb < 0
        ,
        \\
        \abs*{ \log_{\sltparam}\parens{\sltparam / \abs{\imscl}} }
        &
        \text{for}\:
        \abs{\imscl} > \sltparam,\:
        \abs{\perturb\imscl} \leq \sltparam
        ,
        \\
        \abs*{ \log_{\sltparam}\parens{\sltparam / \abs{\perturb \imscl}} }
        &
        \text{for}\:
        \abs{\imscl} \leq \sltparam,\:
        \abs{\perturb\imscl} > \sltparam
        ,
        \\
        0
        &
        \text{otherwise}
        .
    \end{cases}
    \label{eq:mslae-rational-error}
\end{align}

For comparison purposes,
and as it served as inspiration
for the proposed \gls{mslae} loss,
let us also define the \gls{mmuae} loss function as
\begin{align}
    \mmuaeloss\parens{\imvec, \prediction{\imvec}}
    =
    \onenorm{
        \mutransform\parens{\imvec} - \mutransform\parens{\prediction{\imvec}}
    }
    \label{eq:muae}
    ,
\end{align}
where \( \mutransformdef \)
is the \gls{mu-law} transform
(commonly used in audio companding algorithms)
defined element-wise as
\begin{align}
    \mutransform\parens{\imcomponent}
    =
    \sign\parens{\imcomponent}
    \ln\parens*{\frac{1 + \mu \abs{\imcomponent}}{1 + \mu}}
    \label{eq:mut-compression}
    ,
\end{align}
where \( \mu \in \realnumbers_{+} \) defines the extent
of dynamic range compression.
Note that to obtain a dynamic range compression similar to that of
\cref{eq:sup:sl-transform},
\( \mu \) must be set to \( \sltparam^{-1} \).
Proceeding in the same way as for the derivation of
\cref{eq:mslae-rational-error},
and using \( \nu = 1 + \mu \),
one can express the resulting loss for \gls{mmuae} as
\begin{align}
    &
    \mmuaeloss\parens{\imscl , \perturb \imscl}
    =
    \abs{\mutransform \parens{\imscl} - \mutransform\parens{\perturb\imscl}}
    \nonumber
    \\
    &
    \quad
    =
    \begin{cases}
        \abs*{
            \log_{\nu}\parens*{
                \parens{1 + \mu \perturb \abs{\imscl}}
                /
                \parens{1 + \mu \abs{\imscl}}
            }
        }
        &
        \text{for}\:
        \perturb > 0
        ,
        \\
        \abs*{
            \log_{\nu}\parens*{
                \parens{1 - \mu \perturb \abs{\imscl}}
                \parens{1 + \mu \abs{\imscl}}
            }
        }
        &
        \text{for}\:
        \perturb < 0
        .
    \end{cases}
    \label{eq:mmuae-rational-error}
\end{align}

From \cref{eq:mslae-rational-error} and \cref{eq:mmuae-rational-error},
one can note that both losses are not differentiable
for \( \perturb = 1 \),
namely for \( \prediction{\imscl} = \imscl \) (similarly to \gls{mae}).
They both penalize sign errors,
which means that they can preserve the \gls{rf} property of \gls{us} images.
The main advantage of \gls{mslae} over \gls{mmuae} resides in the fact that,
for any true \( \imscl \) and \( \perturb > 0 \) such that
\( \abs{\imscl}, \abs{\perturb \imscl} > \sltparam \),
the loss is a positive constant value
(\gls{ie}, independent of \( \imscl \)).
Consequently, a specific error ratio between a prediction and its true
counterpart is penalized equally, regardless of the true value.
(Note that \gls{mmuae} approximates such a behavior.)
This is a highly desirable feature when working on \gls{hdr} data
such as it is the case in (\gls{rf}) \gls{us} imaging.
Due to the \enquote{threshold} parameter \( \sltparam \),
\gls{mslae} is also not differentiable in a few other cases,
namely for
\( \abs{\imscl} = \sltparam \) and/or \( \abs{\perturb \imscl} = \sltparam \).
Also,
note that in cases where both
\( \abs{\imscl}, \abs{\perturb \imscl} < \sltparam \),
the penalty is zero.
Therefore,
\( \sltparam \) must be selected carefully based on the statistics
of the dataset considered.

The same derivation can be applied to both \gls{mse} and \gls{mae} losses,
resulting in
\begin{align}
    \mseloss\parens{\imscl , \perturb \imscl}
    &=
    \parens{ 1 - \perturb }^{2} \imscl^{2}
    \label{eq:mse-rational-error}
    ,
    \\
    \maeloss\parens{\imscl , \perturb \imscl}
    &=
    \abs{ \parens{1 - \perturb} \imscl }
    \label{eq:mae-rational-error}
    .
\end{align}
From \cref{eq:mse-rational-error,eq:mae-rational-error},
it is clear that \gls{mse} and \gls{mae} are not optimal in the context
of \gls{hdr} data as the resulting loss value is proportional to
the true value \( \imscl \)
(\gls{ie}, quadratically for \gls{mse} and linearly for \gls{mae}).

\section{\sectitleexperiments}%
\label{sec:sup:experiments}

\subsection{Imaging Configurations}%
\label{sec:sup:experiments:imaging-configurations}

\begin{figure*}[t]
    \centering
    \includegraphics{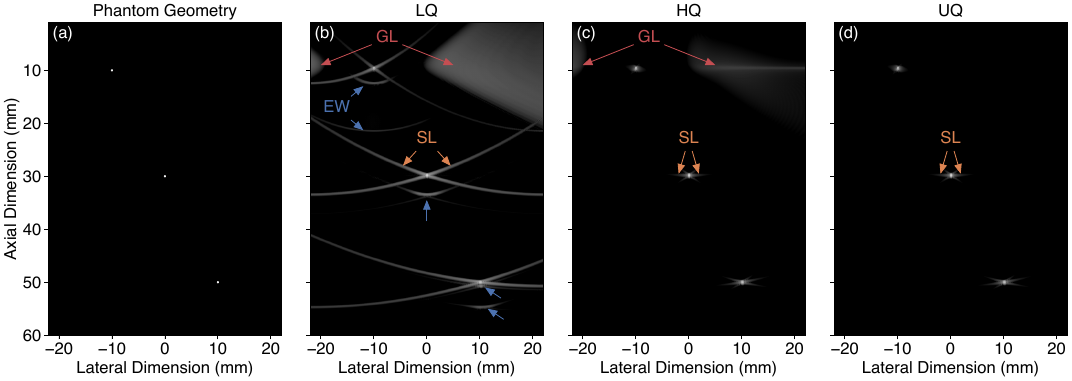}%
    {\phantomsubcaption\label{fig:experiments:psf-example:mask}}%
    {\phantomsubcaption\label{fig:experiments:psf-example:lq}}%
    {\phantomsubcaption\label{fig:experiments:psf-example:hq}}%
    {\phantomsubcaption\label{fig:experiments:psf-example:uq}}%
    \caption{%
        \Gls{bmode} image representations
        (\num{\datasetrangevaldb}-\si{\decibel} range)
        of simulated \glsxtrfull{psf} examples:
        \subref{fig:experiments:psf-example:mask}
        point reflector positions in which the \glsxtrshortpl{psf}
        were evaluated;
        images reconstructed using each imaging configuration considered
        (\cref{tab:imaging-configurations}),
        namely
        \subref{fig:experiments:psf-example:lq}
        \glsxtrfull{lq} configuration,
        \subref{fig:experiments:psf-example:hq}
        \glsxtrfull{hq} configuration
        (\gls{ie}, gold-standard image for the physical transducer array),
        and
        \subref{fig:experiments:psf-example:uq}
        \glsxtrfull{uq} configuration
        (\gls{ie}, gold-standard image for the spatially oversampled virtual
        version of the transducer array, considered as ground-truth).
        Some zones dominated by \glsxtrfull{gl},
        \glsxtrfull{sl},
        and \glsxtrfull{ew} artifacts
        are highlighted by colorized arrows and associated annotations.%
    }%
    \label{fig:experiments:psf-example}
\end{figure*}

As the \gls{psf} of \gls{das}-based pulse-echo \gls{us} imaging systems
is spatially varying,
especially when considering ultrafast acquisitions,
a generic analysis is a complicated task.
Yet,
the \gls{psf} varies slowly over the image domain and its visualization
at some locations in the image provides meaningful information
about its spread and enables comparing different imaging configurations.
\Cref{fig:experiments:psf-example} shows simulated \glspl{psf},
evaluated in three distinct positions,
for the \gls{lq}, \gls{hq}, and \gls{uq}
imaging configurations.
One can note that \gls{gl} artifacts are drastically reduced between
\gls{lq} and \gls{hq},
and are completely removed for \gls{uq}.
Artifacts caused by \glspl{sl} are easily identifiable as they develop
from main lobes in \enquote{cross}-like artifacts.
The spread and amplitude of these artifacts are drastically reduced
for \gls{hq} and \gls{uq} \gls{wrt} \gls{lq}.
Artifacts caused by \glspl{ew},
which are only present in the \gls{lq} configuration
[\cref{fig:experiments:psf-example:lq}],
are the most spatially varying ones and appear as two \enquote{defocused}
duplicates below each main lobe
(except in the center of the lateral dimension where they interfere coherently).
The deeper the position in the image,
the closer \gls{ew} artifacts are to the main lobe,
and the more they resemble the combination of a main lobe
and associated \glspl{sl}.

From these observations,
it is clear that all three imaging configuration considered are characterized
by spatially varying \glspl{psf},
and that this spatially varying property is most pronounced for the \gls{lq}
configuration.
Therefore,
as we seek to learn a restoration mapping (using a \gls{cnn})
to recover high-quality estimates from low-quality ones,
such a mapping needs to be nonstationary as well.
Moreover,
as the \gls{psf} of the \gls{lq} configuration spreads
over a large portion of the image because of diffraction artifacts,
the restoration mapping needs a large receptive field to be effective.
This observation was critical to the design
of the proposed \gls{cnn} architecture
(\cref{sec:methods:network-architecture}).
In particular,
its multiscale structure results in a large receptive field even when using
convolutional kernels of small supports
(\num{3 x 3} in our case).

\section{\exptitletrainings}%
\label{sec:sup:hyperparameters}

Hundreds of training experiments were carried out heuristically
to select the hyperparameters involved in the proposed approach.
The selected training experiments presented in this section
are the ones that guided the selection of the trained \glspl{cnn} evaluated in
\cref{sec:results}.
Each experiment was conducted using the global setup parameters
as well as the training and validation strategy described in
\cref{sec:experiments:training-setup}.
Recall that performances were evaluated on a validation set of
\num{\validsetsize} image pairs (extracted from the simulated dataset)
by computing both the \gls{psnr} and the \gls{ssim} index~\cite{Wang_TIP_2004}
at each validation step (\gls{ie}, every \num{\stepsperepoch} iterations)
on \gls{bmode} representations
(\gls{ie}, log-compressed envelope-detected images)
between
\SIrange[
    range-phrase={ and }
]{\datasetlowervaldb}{\datasetuppervaldb}{\decibel}
(confidence interval detailed in
\cref{sec:sup:methods:rayleigh-statistics:confidence-interval}),
and averaged over the entire validation set.
As the \gls{bmode} \gls{ssim} correlated particularly well
with visual assessments for evaluating the overall quality of recovered images,
it was used to monitor training experiments
and to select the best performing \gls{cnn} instance among
the \num{500} validation steps of each training experiment.

\subsection{\exptitlesignaltype}%
\label{sec:sup:hyperparameters:signal-type}

\Glsxtrlong{us} images can be expressed, analyzed, and displayed
in different representations,
namely \gls{rf}, \gls{iq}, envelope, and \gls{bmode} (log-compressed envelope).
We thus compared the impact of training on these different image representations
using the proposed residual \gls{cnn}
(\cref{fig:methods:network-architecture})
deployed with \num{16} initial expansion channels,
\glspl{res-conv-block},
and additive intrinsic skip connections.
All instances were trained using the \gls{mslae} as loss function
and \gls{uq} images as references,
except when trained on \gls{bmode} representations
in which case the \gls{mae} was used,
as this image representation is already log-compressed.

Even though it may seem intuitive to train
on image representations that we actually look at (\gls{ie}, \gls{bmode}),
it is clear (\cref{fig:results:trainings:signal-reference})
that trainings performed on both envelope and \gls{bmode} representations
are worse than those performed on \gls{rf} and \gls{iq} ones.
This presumably comes from the fact that both
envelope and \gls{bmode} representations
do not contain the \gls{rf} property of \gls{us} images anymore
(due to the envelope detection process),
a property carrying additional information that can be exploited
by the learning process.
The images from \glspl{cnn} trained on \gls{bmode}
and envelope representations are characterized by blurred speckle patterns
[\cref{fig:results:trainings:signal-reference-bmodes:cnnA,%
fig:results:trainings:signal-reference-bmodes:cnnB}].

Trainings performed on \gls{rf} and \gls{iq} representations
resulted in similar performances.
This was expected as \gls{rf} and \gls{iq} images contain the same information.
Hence, both are valid choices.
Yet,
training (and inference) on \gls{iq} representations is more demanding
than on \gls{rf} ones as \gls{iq} images are composed of \enquote{two channels}
(\gls{ie}, real and imaginary parts),
but this only affects the first and last \glspl{conv-layer}.
(\gls{ie}, initial channel expansion and final channel contraction,
\cref{fig:methods:network-architecture}).
On the other hand,
the use of \gls{iq} images simplifies the following envelope detection step
compared with \gls{rf} ones,
namely a simple element-wise modulus compared with a Hilbert transform
(followed by an element-wise modulus).
This is the reason why \gls{iq} was preferred.

\subsection{\exptitlereferenceimage}%
\label{sec:sup:hyperparameters:reference-image}

In our preliminary work~\cite{Perdios_IUS_2018},
we observed that training on reference images
in which \gls{gl} artifacts were still present (\gls{ie}, \gls{hq})
resulted in predicted images with a surprising reduction of said artifacts.
This observation inspired us to develop (and simulate)
reference images free from these artifacts (\gls{ie}, \gls{uq}).
In this experiment,
we evaluated the effect of using \gls{uq} images,
obtained from the optimal (and virtual) \gls{uq} imaging configuration,
as reference images during training,
compared to using \gls{hq} images,
obtained from the \gls{hq} imaging configuration
(\cref{sec:experiments:imaging-configurations}).
In both cases,
\gls{uq} images were used as references for computing validation metrics.
As for \cref{sec:sup:hyperparameters:signal-type},
we used a \gls{cnn} with \num{16} initial expansion channels,
\glspl{res-conv-block}, and additive intrinsic skip connections.
Each instance was trained on \gls{iq} representations using the \gls{mslae}
as loss function.

\begin{figure}[t]
    \centering
    \includegraphics{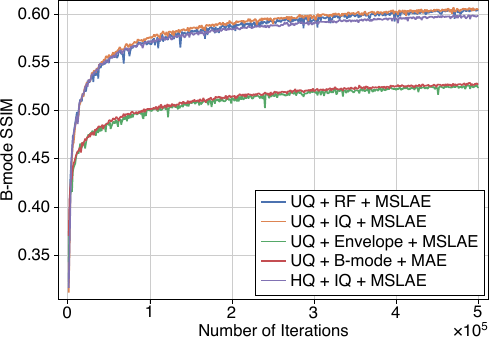}%
    \caption{
        Validation metric curves
        (\glsxtrshort{ssim} evaluated on \gls{bmode} representations)
        of training experiments performed using
        different image representations
        (\gls{ie}, \glsxtrshort{rf}, \glsxtrshort{iq}, envelope,
        and \glsxtrshort{bmode})
        and different reference images
        (\gls{ie}, \glsxtrshort{hq} and \glsxtrshort{uq}).
        All training experiments were performed on identical instances of
        the proposed residual \glsxtrshort{cnn}
        (\cref{fig:methods:network-architecture})
        with \num{16} initial expansion channels,
        \glsxtrshortpl{res-conv-block},
        and additive intrinsic skip connections.%
    }%
    \label{fig:results:trainings:signal-reference}
\end{figure}

\begin{figure*}[t]
    \centering
    \includegraphics{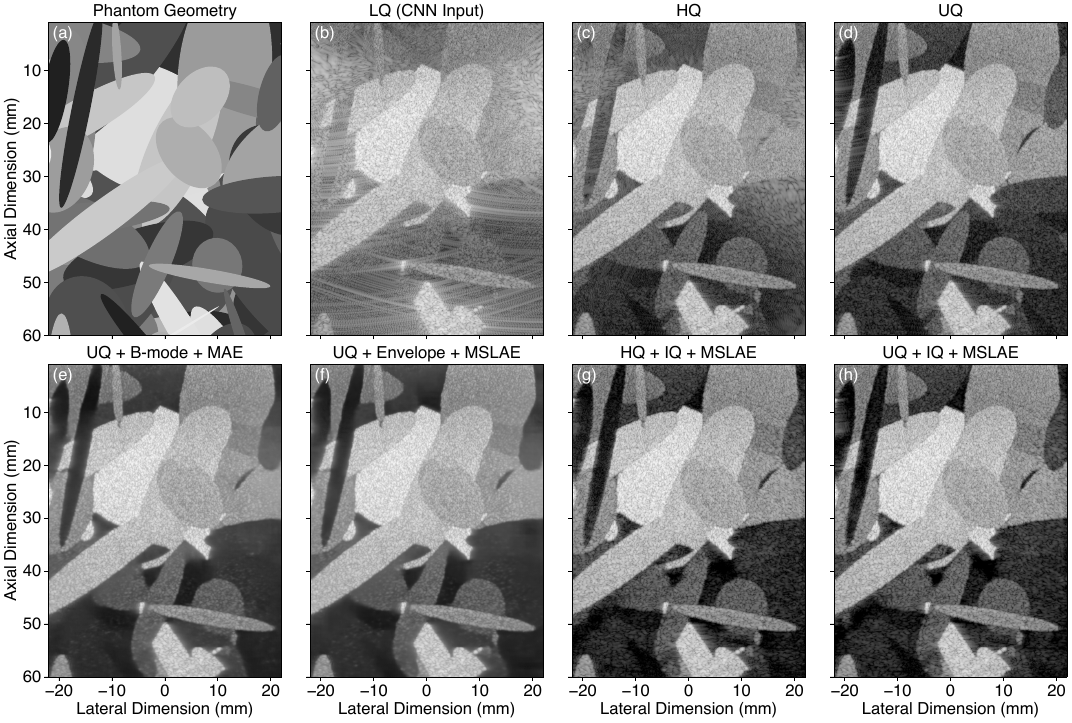}%
    {\phantomsubcaption\label{fig:results:trainings:signal-reference-bmodes:mask}}%
    {\phantomsubcaption\label{fig:results:trainings:signal-reference-bmodes:lq}}%
    {\phantomsubcaption\label{fig:results:trainings:signal-reference-bmodes:hq}}%
    {\phantomsubcaption\label{fig:results:trainings:signal-reference-bmodes:uq}}%
    {\phantomsubcaption\label{fig:results:trainings:signal-reference-bmodes:cnnA}}%
    {\phantomsubcaption\label{fig:results:trainings:signal-reference-bmodes:cnnB}}%
    {\phantomsubcaption\label{fig:results:trainings:signal-reference-bmodes:cnnC}}%
    {\phantomsubcaption\label{fig:results:trainings:signal-reference-bmodes:cnnD}}%
    \caption{%
        \Gls{bmode} image representations
        (\num{\datasetrangevaldb}-\si{\decibel} range)
        of a numerical test phantom sample
        (extracted from the simulated dataset):
        \subref{fig:results:trainings:signal-reference-bmodes:mask}
        the phantom mask;
        images reconstructed using each imaging configuration considered
        (\cref{tab:imaging-configurations}),
        namely
        \subref{fig:results:trainings:signal-reference-bmodes:lq}
        \glsxtrfull{lq},
        \subref{fig:results:trainings:signal-reference-bmodes:hq}
        \glsxtrfull{hq}
        (\gls{ie}, gold-standard image for the physical transducer array),
        and
        \subref{fig:results:trainings:signal-reference-bmodes:uq}
        \glsxtrfull{uq}
        (\gls{ie}, reference image);
        images recovered from the \glsxtrshort{lq} input image
        using the proposed approach with different \glsxtrfullpl{cnn},
        deployed with
        \num{16} initial expansion channels,
        \glsxtrfullpl{res-conv-block},
        and additive intrinsic skip connections,
        trained on different image representations
        and image references,
        namely
        \subref{fig:results:trainings:signal-reference-bmodes:cnnA}
        \glsxtrshort{uq} + \gls{bmode} + \glsxtrfull{mae},
        \subref{fig:results:trainings:signal-reference-bmodes:cnnB}
        \glsxtrshort{uq} + envelope + \glsxtrfull{mslae},
        \subref{fig:results:trainings:signal-reference-bmodes:cnnC}
        \glsxtrshort{hq} + \glsxtrfull{iq} + \glsxtrshort{mslae},
        and
        \subref{fig:results:trainings:signal-reference-bmodes:cnnD}
        \glsxtrshort{uq} + \glsxtrshort{iq} + \glsxtrshort{mslae}.%
    }%
    \label{fig:results:trainings:signal-reference-bmodes}
\end{figure*}

\Cref{fig:results:trainings:signal-reference} demonstrates
the benefit of training on \gls{uq} rather than \gls{hq} reference images
in terms of \gls{bmode} \gls{ssim}.
As expected, we observed that imaging artifacts,
in particular those caused by \glspl{gl},
were better reduced when trainings were performed
using \gls{uq} images as references
[\cref{fig:results:trainings:signal-reference-bmodes:cnnD}],
than when using \gls{hq} ones
[\cref{fig:results:trainings:signal-reference-bmodes:cnnC}].
Details initially hidden by \gls{gl} artifacts were also better recovered.
Yet,
and as observed in our preliminary work~\cite{Perdios_IUS_2018},
the training on \gls{hq} images as references resulted in images
with far less \gls{gl} artifacts compared with the corresponding
\gls{hq} images
[\cref{fig:results:trainings:signal-reference-bmodes:hq,%
fig:results:trainings:signal-reference-bmodes:cnnC}].
This effect remains unexplained and could therefore be unpredictable.
Nonetheless,
training \glspl{cnn} on the newly designed \gls{uq} reference images
resulted in a more consistent \gls{gl} reduction
as \gls{uq} reference images did not contain such artifacts.
Therefore,
we opted for using \gls{uq} images as reference images during training.

\subsection{\exptitleloss}%
\label{sec:sup:hyperparameters:loss}

We compared the effect of using different training losses,
namely the \gls{mse},
the \gls{mae},
and the proposed \gls{mslae}
(implemented with a \enquote{threshold} parameter \( \sltparam \)
corresponding to \SI{\datasetlowervaldb}{\decibel}).
For this comparison,
we used a \gls{cnn} with \num{16} initial expansion channels,
\glspl{res-conv-block}, and additive intrinsic skip connections.
Trainings were performed on \gls{iq} representations using \gls{uq} images
as references.

\begin{figure}[t]
    \centering
    \includegraphics{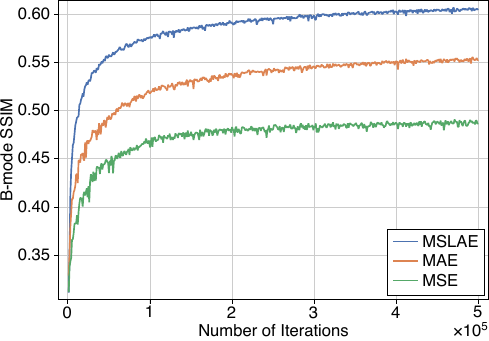}%
    \caption{
        Validation metric curves
        (\glsxtrshort{ssim} evaluated on \gls{bmode} representations)
        of training experiments performed using
        different training losses
        (\gls{ie}, \glsxtrshort{mse}, \glsxtrshort{mae},
        and \glsxtrshort{mslae}).
        All training experiments were performed on identical instances of
        the proposed residual \glsxtrshort{cnn}
        (\cref{fig:methods:network-architecture})
        with \num{16} initial expansion channels,
        \glsxtrshortpl{res-conv-block},
        and additive intrinsic skip connections.%
    }%
    \label{fig:results:trainings:loss-comparisons}
\end{figure}

\Cref{fig:results:trainings:loss-comparisons} clearly shows that,
despite being the standard loss in regression problems,
and the loss we used in our preliminary work~\cite{Perdios_IUS_2018},
the \gls{mse} is the least effective one to address the restoration
problem involved in the proposed approach.
Indeed, the \gls{hdr} property of \gls{rf} \gls{us} images makes the use of
the \gls{mse} suboptimal,
as too much emphasis is put on image samples with
large values (\gls{ie}, highly echogenic).
The use of the \gls{mae} as loss function,
which has been increasingly reported in similar regression problems such as
image super-resolution~\cite{Lim_CVPRW_2017} and
\gls{mri}~\cite{Zhao_TIP_2019},
performed better than using the \gls{mse}.
As expected,
the fact that \gls{mae} is less sensitive to \enquote{outliers}
makes it more robust to \gls{hdr} contents.
A substantial increase in performance \gls{wrt} \gls{mae} was observed
when using the proposed \gls{mslae} as loss function,
confirming its superiority over the other two losses compared
for learning a restoration mapping on \gls{hdr} \gls{rf} (or \gls{iq}) data.

One can also note that, the more effective the loss,
the more stable the training,
appearing as smoother validation curves.
It was also observed that the use of a more effective loss resulted in
trainings less prone to overfitting,
thus less demanding in terms of data quantity,
as it maximized the usage of the available information content.
The flattening of the validation curve observed when using the \gls{mse}
as loss function is an indication
that overfitting would most probably appear earlier than when using
the other losses.
Further analysis and discussions on losses can be found in
\cref{sec:results,sec:discussion}.

\subsection{\exptitlearchitecture}%
\label{sec:sup:hyperparameters:architecture}

These experiments were conducted to evaluate the effects of the
proposed \gls{cnn} architecture improvements
(\cref{sec:methods:network-architecture}).
All trainings were performed on \gls{iq} representations
using \gls{mslae} as loss function
and \gls{uq} images as references.
All experiments were carried out on \gls{cnn} instances
with \num{16} initial expansion channels.
Two types of intrinsic skip connections,
namely additive and concatenated as originally proposed
in~\cite{Ronneberger_MICCAI_2015},
were compared
on \gls{cnn} instances with traditional \glspl{std-conv-block}.
We also compared the use of the proposed \glspl{res-conv-block} instead of
\glspl{std-conv-block} on \gls{cnn} instances
with additive intrinsic skip connections,
as concatenated ones cannot be used with \glspl{res-conv-block} directly.

\begin{figure}[t]
    \centering
    \includegraphics{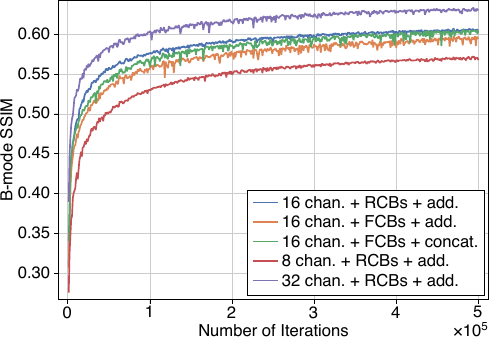}%
    \caption{
        Validation metric curves
        (\glsxtrshort{ssim} evaluated on \gls{bmode} representations)
        of training experiments performed using
        different combinations of
        initial channel expansion numbers
        (\gls{ie}, \num{8}, \num{16}, and \num{32}),
        convolutional blocks (\gls{ie}, \glsxtrshortpl{res-conv-block}
        and \glsxtrshortpl{std-conv-block}),
        and intrinsic skip connections (\gls{ie}, additive and concatenated).
        Each training experiment was performed
        using \glsxtrshort{mslae} as loss function.%
    }%
    \label{fig:results:trainings:architecture-comparisons}
\end{figure}

The comparison of concatenated and additive intrinsic skip connections
implemented with the jointly compatible \glspl{std-conv-block}
shows (\cref{fig:results:trainings:architecture-comparisons}) that
the use of concatenated ones
results in slightly better performances than additive ones.
This was somehow expected as the use of concatenated intrinsic skip connections
increases the total number of trainable parameters
(\gls{ie}, increased capacity)
by approximately \SI{7}{\percent} in the \enquote{decoding} arm only
[\cref{fig:methods:network-architecture:main}].
As a result it also significantly increases both training and inference times,
due to augmented convolution operations which are the most costly ones.
(Especially the last intrinsic skip connection which results in the most
computationally intense convolutional operation of the \gls{cnn} architecture.)
The comparison of \glspl{res-conv-block} and \glspl{std-conv-block}
implemented with additive intrinsic skip connections
showed that the use of \glspl{res-conv-block} performs best
at same \gls{cnn} capacity
with (almost) no effect on the inference time.
It also outperformed the greater-capacity \gls{cnn} instance with
\glspl{std-conv-block} and concatenated intrinsic skip connections
both in terms of validation metric (\gls{bmode} \gls{ssim})
and training stability (smoother validation curve).
The use of \glspl{res-conv-block} together with
additive intrinsic skip connections was therefore selected.

\subsection{\exptitlechannelnumber}%
\label{sec:sup:hyperparameters:channel-number}

Since the initial channel expansion number affects the entire architecture,
this parameter has a major impact on
the overall \gls{cnn} capacity,
the training time,
and the inference time.
Three \gls{cnn} instances
(\glspl{res-conv-block}, additive intrinsic skip connections)
with initial channel expansion numbers of \numlist{8;16;32}
were trained on \gls{iq} representations
using \gls{mslae} as loss function
and \gls{uq} images as references.
In these settings,
the total number of trainable parameters were
\num{687720}, \num{2748624}, and \num{10989984}, respectively
(\gls{ie}, approximately quadrupled when the initial channel expansion number
is doubled).

As \gls{us} imaging is,
in essence,
a real-time imaging modality,
inference speed tests were also performed on these three configurations.
To quantify the impact on the inference time of using \gls{iq} images
rather than \gls{rf} images
(\cref{sec:sup:hyperparameters:reference-image}),
inference speed tests were also performed on the same configurations
but trained on \gls{rf} images.
We computed the averaged inference time, over \num{5000} runs,
on images of size \( \numimdim \), with appropriate zero-padding,
for each configuration using both
\gls{tensorflow}\footnote{\tfurl} (\tfver)
and
\gls{tensorrt}\footnote{\trturl} (\trtver), an inference optimizer.
Different \gls{gpu} models were compared, namely
the \gls{mx150} (laptop, \num{384} cores, Pascal arch., entry-level),
the \gls{1080ti} (desktop, \num{3584} cores, Pascal arch.),
and the \gls{titanv} (desktop, \num{5120} cores, Volta arch.).

As expected,
the more initial expansion channels
the better the validation metric
(\cref{fig:results:trainings:architecture-comparisons}),
provided that enough data is available to avoid overfitting.
Inference speed tests
(\cref{tab:results:trainings:inference-timings}) demonstrated that,
depending on code optimization and \gls{gpu} model,
real-time imaging is feasible using the proposed approach
and a \num{16}-channel version.
Since we are using simulations and can theoretically generate
a dataset of infinite size preventing from any overfitting,
the architecture optimization really comes down to speed \gls{vs} quality
in scenarios where real-time imaging is a necessity.
One can also note that the increase in inference time of using \gls{iq} images
rather than \gls{rf} images was of about \SIrange{5}{10}{\percent},
and did not result in loosing real-time capabilities.

\begin{table}[t]
    \tablefontstyle
    \centering
    \caption{%
        Average Inference Time for Different Image Representations
        and Initial Channel Expansion Numbers%
    }%
    \label{tab:results:trainings:inference-timings}%
    
\sisetup{
    table-figures-integer = 3,
    table-figures-decimal = 0,
    table-number-alignment = center,
    table-space-text-post = \,\si{\milli\second}
}

\newcommand*{\tnotetfmark}{a}
\newcommand*{\tnotetrtmark}{b}
\newcommand*{\tnotememmark}{c}

\newcommand{\specialcell}[2][c]{%
  \begin{tabular}[#1]{@{}c@{}}#2\end{tabular}}

\begin{threeparttable}
\begin{tabular}{
    c
    S[table-figures-integer = 2, table-space-text-post]
    S[table-figures-integer = 3]
    S[table-figures-integer = 3]
    S[table-figures-integer = 2]
    S[table-figures-integer = 2]
    S[table-figures-integer = 2]
    S[table-figures-integer = 2]
}
    \toprule
    \mco{\multirow{2}{*}{\specialcell{\tableheaderstyle{Image}\\\tableheaderstyle{Repr.}}}}
    & \mco{\multirow{2}{*}{\specialcell{\tableheaderstyle{Channel}\\\tableheaderstyle{Number}}}}
    & \multicolumn{2}{c}{\tableheaderstyle{MX150}}
    & \multicolumn{2}{c}{\tableheaderstyle{1080 Ti}}
    & \multicolumn{2}{c}{\tableheaderstyle{TITAN V}}
    \\
    &
    & \mco{TF\tnote{\tnotetfmark}} & \mco{TRT\tnote{\tnotetrtmark}}
    & \mco{TF} & \mco{TRT}
    & \mco{TF} & \mco{TRT}
    \\
    \midrule
    \multirow{3}{*}{\glsxtrshort{rf}}
    & 8
    & 130\,\si{\milli\second} &  83\,\si{\milli\second}
    &  21\,\si{\milli\second} &  10\,\si{\milli\second}
    &  18\,\si{\milli\second} &  8\,\si{\milli\second}
    \\
    & 16
    & 249\,\si{\milli\second} & 167\,\si{\milli\second}
    &  36\,\si{\milli\second} &  21\,\si{\milli\second}
    &  29\,\si{\milli\second} &  14\,\si{\milli\second}
    \\
    & 32
    & \mco{\invalidtabentry\tnote{\tnotememmark}}
    & \mco{\invalidtabentry\tnote{\tnotememmark}}
    &  74\,\si{\milli\second} &  52\,\si{\milli\second}
    &  52\,\si{\milli\second} &  37\,\si{\milli\second}
    \\
    \midrule
    \multirow{3}{*}{\glsxtrshort{iq}}
    & 8
    & 136\,\si{\milli\second} &  86\,\si{\milli\second}
    &  24\,\si{\milli\second} &  12\,\si{\milli\second}
    &  21\,\si{\milli\second} &  9\,\si{\milli\second}
    \\
    & 16
    & 256\,\si{\milli\second} & 172\,\si{\milli\second}
    &  39\,\si{\milli\second} &  22\,\si{\milli\second}
    &  32\,\si{\milli\second} &  15\,\si{\milli\second}
    \\
    & 32
    & \mco{\invalidtabentry\tnote{\tnotememmark}}
    & \mco{\invalidtabentry\tnote{\tnotememmark}}
    &  77\,\si{\milli\second} &  53\,\si{\milli\second}
    &  54\,\si{\milli\second} &  39\,\si{\milli\second}
    \\
    \bottomrule
\end{tabular}
\begin{tablenotes}[para]
    \item[\tnotetfmark] \gls{tensorflow}
    \item[\tnotetrtmark] \gls{tensorrt}
    \item[\tnotememmark] Not enough memory
\end{tablenotes}
\end{threeparttable}%
\end{table}

\subsection{\exptitletrainingsize}%
\label{sec:sup:hyperparameters:training-size}

This experiment was performed to evaluate the impact of the training set
size, and most importantly,
to guarantee that the selected configuration is not prone to overfitting.
We considered
the proposed residual \gls{cnn} (\cref{fig:methods:network-architecture})
deployed with \num{16} initial expansion channels,
\glspl{res-conv-block},
and additive intrinsic skip connections.
Each instance was trained on \gls{iq} representations
using \gls{mslae} as loss function
and \gls{uq} images as references.
Different training set sizes (spanning a logarithmic range) were compared,
namely \numlist{200;409;837;1713;3504;7168;14664;30000}.

\begin{figure}[t]
    \centering
    \includegraphics{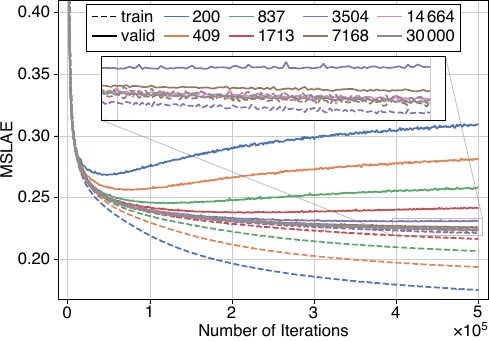}%
    \caption{
        Training and validation loss curves
        of training experiments performed using
        different training set sizes.
        All training experiments were performed on identical instances of
        the proposed residual \glsxtrshort{cnn}
        (\cref{fig:methods:network-architecture})
        with \num{16} initial expansion channels,
        \glsxtrshortpl{res-conv-block},
        and additive intrinsic skip connections,
        using \glsxtrshort{mslae} as loss function.%
    }%
    \label{fig:results:trainings:training-size-comparisons}
\end{figure}

From \cref{fig:results:trainings:training-size-comparisons},
it is evident that the training of the analyzed,
comparatively small \gls{cnn} with only \num{16} initial expansion channels,
suffers from obvious overfitting up to \num{\sim{} 7}\si{\kilo} training
image pairs.
In these settings,
it seems like the use of a training set composed of \num{\sim{} 10}\si{\kilo}
image pairs would be sufficient to avoid overfitting.
Yet, the training of \glspl{cnn} with more capacity,
such as with \num{32} initial expansion channels,
necessarily requires larger training sets.
Thus,
we chose to use \num{\trainsetsize} image pairs for the reported experiments.
One can also note
(magnified inset of \cref{fig:results:trainings:training-size-comparisons})
that the larger the training set,
the better the resulting validation loss,
even after \enquote{obvious} overfitting cases.

\subsection{\exptitlekernelinitializer}%
\label{sec:sup:hyperparameters:kernel-initializers}

We confirmed our choice of using \gls{glorot} uniform as kernel initializer
by comparing the performances of differently initialized \glspl{cnn} instances
with \num{16} initial expansion channels,
\glspl{res-conv-block}, and additive intrinsic skip connections.
As the proposed architecture is composed of \glspl{conv-layer}
and \gls{relu} activations,
we were particularly interested in evaluating the \gls{he} initializer
proposed in~\cite{He_ICCV_2015} to maintain the variance through such layers
and activations.
We compared both \gls{glorot}~\cite{Glorot_AISTATS_2010} and \gls{he} initializers
implemented with uniform and normal distributions.
All kernels were initialized by the initializers considered,
except for the initial channel expansion layer
and the final channel contraction layer
which were always initialized using the \gls{glorot} (uniform) initializer,
as they are not followed by a \gls{relu} activation.

\begin{figure}[t]
    \centering
    \includegraphics{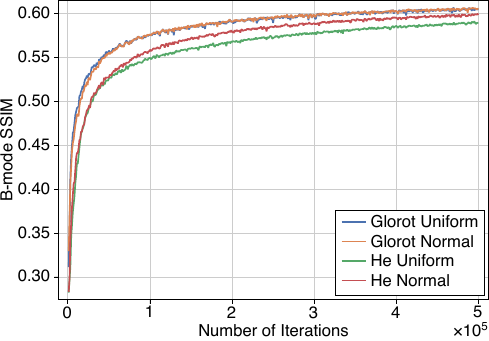}%
    \caption{
        Validation metric curves
        (\glsxtrshort{ssim} evaluated on \gls{bmode} representations)
        of training experiments performed using
        different kernel initializers
        (\gls{ie}, \gls{glorot} uniform, \gls{glorot} normal,
        \gls{he} uniform, and \gls{he} normal).
        Each training experiment was performed on
        the proposed residual \gls{cnn}
        (\cref{fig:methods:network-architecture})
        with \num{16} initial expansion channels,
        \glsxtrshortpl{res-conv-block},
        and additive intrinsic skip connections,
        using \glsxtrshort{mslae} as loss function.%
    }%
    \label{fig:results:trainings:kernel-initializer-comparisons}
\end{figure}

Interestingly,
both implementations of the \gls{glorot} initializer (\gls{ie}, uniform and normal)
performed similarly better than both implementations of the \gls{he} initializer
(\cref{fig:results:trainings:kernel-initializer-comparisons}).
This may be explained by the many residual connections
(\gls{ie}, all intrinsic ones and the outer one)
and/or the multiscale property
of the proposed architecture,
for which the benefit of \gls{he} initializer does not seem to be confirmed.

\subsection{\exptitlelearningrates}%
\label{sec:sup:hyperparameters:learning-rates}

We also compared different learning rates of \numlist{1e-5;5e-5;1e-4;5e-4;1e-3}.
Identical instances of the proposed residual \gls{cnn}
(\cref{fig:methods:network-architecture})
deployed with \num{16} initial expansion channels,
\glspl{res-conv-block},
and additive intrinsic skip connections
were trained using the Adam optimizer~\cite{Kingma_ARXIV_2014}
with each learning rate.
Trainings were performed using the \gls{mslae} as loss function
and \gls{uq} images as references.

\begin{figure}[t]
    \centering
    \includegraphics{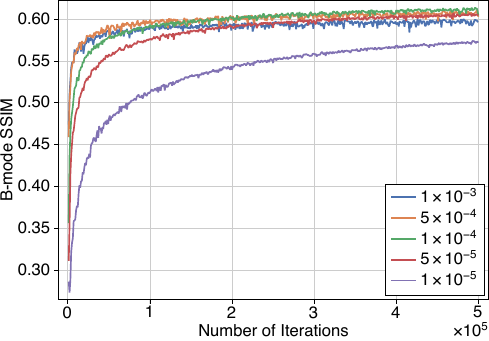}%
    \caption{
        Validation metric curves
        (\glsxtrshort{ssim} evaluated on \gls{bmode} representations)
        of training experiments performed using
        different learning rates
        (\gls{ie}, \numlist{1e-5;5e-5;1e-4;5e-4;1e-3}).
        Each training experiment was performed on
        the proposed residual \glsxtrshort{cnn}
        (\cref{fig:methods:network-architecture})
        with \num{16} initial expansion channels,
        \glsxtrshortpl{res-conv-block},
        and additive intrinsic skip connections,
        using \glsxtrshort{mslae} as loss function.%
    }%
    \label{fig:results:trainings:learning-rate-comparisons}
\end{figure}

From \cref{fig:results:trainings:learning-rate-comparisons},
it is clear that a learning rate of \num{1e-5} is too small
and that a learning rate of \num{1e-3} is too large.
The other three,
namely \numlist{5e-5;1e-4;5e-4},
resulted in fairly similar performances.
Even though a learning rate of \num{5e-5} was the least performing
among these three,
we decided to select this one as it resulted in the most stable validation curve
and adapted best to all other experiments carried out for hyperparameter search
(in particular when training \glspl{cnn} with larger capacities).

\subsection{Summary}%
\label{sec:sup:hyperparameters:summary}

All proposed improvements to the neural network architecture
(\cref{sec:methods:network-architecture})
resulted in increased performances.
The use of optimal (virtual) \gls{uq} images as reference images
for the training process was successful.
It provided better results than using \gls{hq} images,
with controlled \gls{gl} artifacts removal
(\cref{sec:sup:hyperparameters:reference-image}).
The image representation onto which the training is performed is crucial.
Trainings performed on \Gls{bmode} and envelope representations
resulted in much worse image restoration capabilities
than \gls{rf} and \gls{iq} ones,
as the envelope detection process removes high-frequency content
that can be exploited by the \gls{cnn}.
Even though both \gls{iq} and \gls{rf} trainings performed similarly,
as the information contained in both representations is identical,
we opted for the \gls{iq} ones as it allows for a simplified subsequent
envelope detection process
(\cref{sec:sup:hyperparameters:signal-type}).
The loss choice was observed as the most impactful parameter
(\cref{sec:sup:hyperparameters:loss})
and was therefore further evaluated in an \gls{us}-specific test environment
(\cref{sec:experiments:numerical-evaluations,%
sec:results:numerical-evaluations}).

\section{Results}%
\label{sec:sup:results}

\subsection{\exptitlenumerical}%
\label{sec:sup:results:numerical-evaluations}

As we generated a simulated test set obtained
from \num{300} statistically independent realizations
(\gls{ie}, random scatterers)
of the same numerical test phantom
(\cref{sec:experiments:numerical-evaluations}),
we also analyzed the incoherent average
(performed after envelope detection)
of all images reconstructed using the \gls{lq}, \gls{hq}, and \gls{uq}
imaging configurations,
as well as using the proposed approach
with the four trained \glspl{cnn} considered
(\gls{ie}, \resnumcnnAlgd{}, \resnumcnnBlgd{}, \resnumcnnClgd{},
and \resnumcnnDlgd{} defined in \cref{sec:experiments:numerical-evaluations}).
As independent realizations of scatterers
with identical statistical properties
result in similar images with uncorrelated speckle patterns,
the incoherent averaging of a large amount of such images
provides us with an interesting visualization of stationary structures;
the underlying phantom mask and the image zones suffering from imaging artifacts
are fully revealed.

\begin{figure*}[t]
    \centering
    \includegraphics{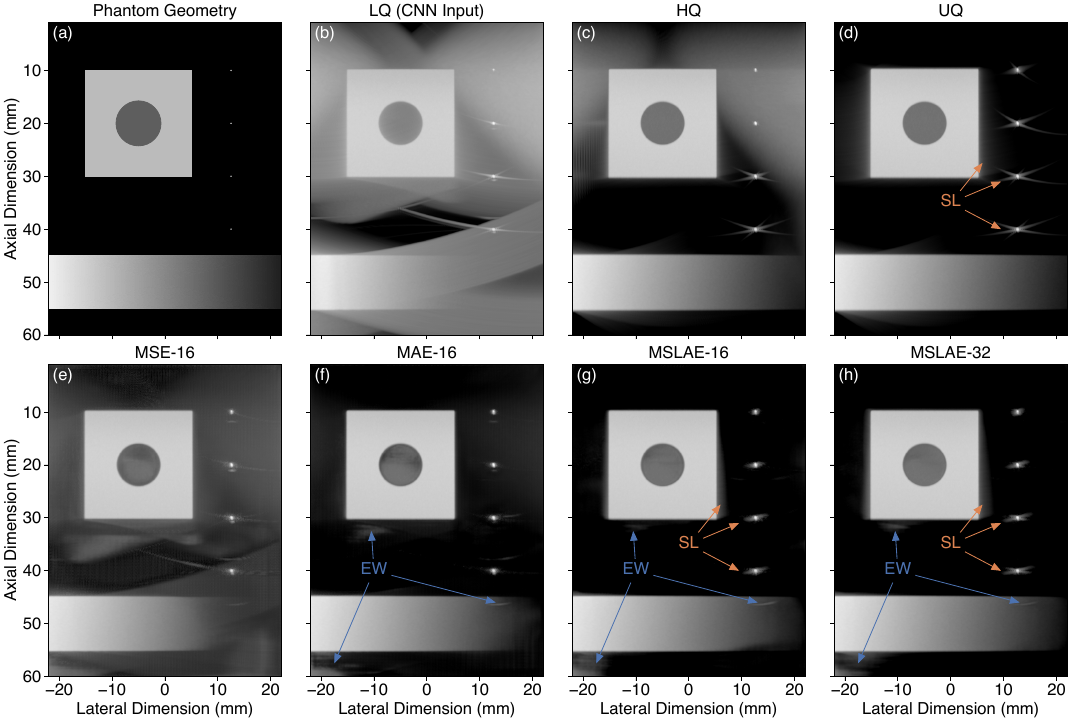}%
    {\phantomsubcaption\label{fig:results:numerical-phantom-mean:mask}}%
    {\phantomsubcaption\label{fig:results:numerical-phantom-mean:lq}}%
    {\phantomsubcaption\label{fig:results:numerical-phantom-mean:hq}}%
    {\phantomsubcaption\label{fig:results:numerical-phantom-mean:uq}}%
    {\phantomsubcaption\label{fig:results:numerical-phantom-mean:cnnA}}%
    {\phantomsubcaption\label{fig:results:numerical-phantom-mean:cnnB}}%
    {\phantomsubcaption\label{fig:results:numerical-phantom-mean:cnnC}}%
    {\phantomsubcaption\label{fig:results:numerical-phantom-mean:cnnD}}%
    \caption{%
        \Gls{bmode} image representations
        (\num{\datasetrangevaldb}-\si{\decibel} range)
        of the incoherent average (performed after envelope detection)
        of all images reconstructed from the \num{\numtestsetsize}
        independent realizations (random scatterers)
        of the numerical test phantom:
        \subref{fig:results:numerical-phantom-mean:mask}
        the phantom mask;
        images reconstructed using each imaging configuration considered
        (\cref{tab:imaging-configurations}),
        namely
        \subref{fig:results:numerical-phantom-mean:lq}
        \glsxtrfull{lq} configuration,
        \subref{fig:results:numerical-phantom-mean:hq}
        \glsxtrfull{hq} configuration
        (\gls{ie}, gold-standard image for the physical transducer array),
        and
        \subref{fig:results:numerical-phantom-mean:uq}
        \glsxtrfull{uq} configuration
        (\gls{ie}, reference image);
        images recovered from the \glsxtrfull{lq} input image
        using the proposed approach
        with each of the trained \glsxtrfullpl{cnn} considered
        (\cref{sec:experiments:numerical-evaluations}),
        namely
        \subref{fig:results:numerical-phantom-mean:cnnA}
        \resnumcnnAlgd{},
        \subref{fig:results:numerical-phantom-mean:cnnB}
        \resnumcnnBlgd{},
        \subref{fig:results:numerical-phantom-mean:cnnC}
        \resnumcnnClgd{},
        and
        \subref{fig:results:numerical-phantom-mean:cnnD}
        \resnumcnnDlgd{}.
        Some remaining \glsxtrfull{sl} and \glsxtrfull{ew} artifacts
        are highlighted by colorized arrows and associated annotations.
    }%
    \label{fig:results:numerical-phantom-mean}
\end{figure*}

The visual assessment of such a representation
(\cref{fig:results:numerical-phantom-mean})
for each image reconstruction method
compared in \cref{sec:results:numerical-evaluations}
leads to the same conclusions,
some of which deserve to be re-emphasized.
The comparison of the averaged restoration of the low-echogenic inclusion
is of particular interest
and shows again the benefit of using the proposed \gls{mslae} as loss function
over the conventional \gls{mse} and \gls{mae} losses
[\cref{fig:results:numerical-phantom-mean:cnnA,%
fig:results:numerical-phantom-mean:cnnB,%
fig:results:numerical-phantom-mean:cnnC}].
By comparing the results obtained with \resnumcnnClgd{} and \resnumcnnDlgd{},
one can note that the greater the \gls{cnn} capacity,
the closer the recovered image to the corresponding \gls{uq} reference.
This increase in performance is especially visible on the remaining
\gls{sl} artifacts,
which more closely resemble those of the \gls{uq} reference.
This visualization makes it very clear that \gls{ew} artifacts are the most
complex to deal with.
It also reveals a remaining \gls{ew} artifact arising from the deepest
bright reflector and located within the log-linear gradient
that was indistinguishable in the test phantom sample
displayed in \cref{fig:results:numerical-phantom}.

\end{document}